\documentclass[aps,prl,twocolumn,showpacs,floatfix,superscriptaddress,noshowpacs]{revtex4-1}
\usepackage[dvips]{graphicx}
\usepackage{amsmath}
\usepackage{amssymb}
\usepackage{color}
\usepackage{float}
\usepackage[colorlinks=true,citecolor=blue,linkcolor=blue]{hyperref}

\begin{document}

\title{Superconductivity in a disordered metal with Coulomb interactions}

\author{Svetlana V. Postolova}
\affiliation{A.V. Rzhanov Institute of Semiconductor Physics, 630090 Novosibirsk, Russia}

\author{Alexey Yu. Mironov}
\affiliation{A.V. Rzhanov Institute of Semiconductor Physics, 630090 Novosibirsk, Russia}

\author{V\'ictor Barrena}
\affiliation{Laboratorio de Bajas Temperaturas y Altos Campos Magn\'eticos (Unidad Asociada UAM-CSIC), Departamento de F\'isica de la Materia Condensada, Instituto Nicol\'as Cabrera and Condensed Matter Physics Center (IFIMAC), Universidad Aut\'onoma de Madrid, E-28049 Madrid,
Spain}

\author{Jose Benito-Llorens}
\affiliation{Laboratorio de Bajas Temperaturas y Altos Campos Magn\'eticos (Unidad Asociada UAM-CSIC), Departamento de F\'isica de la Materia Condensada, Instituto Nicol\'as Cabrera and Condensed Matter Physics Center (IFIMAC), Universidad Aut\'onoma de Madrid, E-28049 Madrid,
Spain}

\author{Jose Gabriel Rodrigo}
\affiliation{Laboratorio de Bajas Temperaturas y Altos Campos Magn\'eticos (Unidad Asociada UAM-CSIC), Departamento de F\'isica de la Materia Condensada, Instituto Nicol\'as Cabrera and Condensed Matter Physics Center (IFIMAC), Universidad Aut\'onoma de Madrid, E-28049 Madrid,
Spain}

\author{Hermann Suderow}
\affiliation{Laboratorio de Bajas Temperaturas y Altos Campos Magn\'eticos (Unidad Asociada UAM-CSIC), Departamento de F\'isica de la Materia Condensada, Instituto Nicol\'as Cabrera and Condensed Matter Physics Center (IFIMAC), Universidad Aut\'onoma de Madrid, E-28049 Madrid,
Spain}

\author{Mikhail R. Baklanov}
\affiliation{North China University of Technology, Beijing 100144, China}

\author{Tatyana I. Baturina}
\affiliation{A.V. Rzhanov Institute of Semiconductor Physics, 630090 Novosibirsk, Russia}

\author{Valerii M. Vinokur}
\affiliation{Materials Science Division, Argonne National Laboratory, Argonne, Illinois 60637, USA}

\begin{abstract}
{We study the electronic densities of states (DOS) of strongly disordered superconducting thin films of TiN. We find, using Scanning Tunneling Microscopy (STM) that the DOS decreases towards the Fermi level in the normal phase obtained by applying magnetic fields. The DOS shows spatial fluctuations whose length scale is related to the energy dependent DOS and is similar in normal and superconducting phases. This suggests that Coulomb interactions lead to a spatially varying DOS in the normal phase of a disordered superconductor.}
\end{abstract}

\maketitle

Superconductor-insulator transition (SIT) is an exemplary quantum phase transition occuring in disordered superconducting films and Josephson junction arrays at low temperatures\,\cite{GoldmanReview,LIN2015130,Vinokur2008}. The SIT can be driven by varying disorder and suppressing electron diffusion and/or by applying a magnetic field and manifests a dramatic change in the ground state of the system from superconducting (zero resistance) to insulator (infinite resistance). The insulator has a self-induced electronic granular structure\,\cite{Zvi} comprising superconducting droplets hosting a superconducting gap\,\cite{TiNInhom, Shahar} and connected by weak links being thus similar to Josephson junction arrays. The superconducting gap does not show coherence peaks, suggesting that Cooper pairs are localized\,\cite{NoPeak}. There is a pseudogap above the superconducting critical temperature $T_c$, suggesting enhanced superconducting fluctuations~\cite{TiNPG}, and the superconducting droplets survive up to temperatures of two or three times $T_c$~\cite{PratapPRL11, PratapScirep13}. All this suggests that Coulomb screening is suppressed and the SIT is driven by the ratio of the characteristic Coulomb and superconducting coupling energies. The suppressed Coulomb screening results in the depletion of the normal state density of states (DOS) near the Fermi level termed the zero bias anomaly (ZBA)\,\cite{AA,Valles89}. In disordered Be films, the DOS evolves upon cooling from a Coulomb-like gap~\cite{Butko2000} to a well opened, semiconducting like gap which is often termed a hard gap~\cite{BeDOS}.

These experiments mostly address the properties of systems with very strong disorder. On the other hand, Bartosch and Kopietz showed that, as a consequence of reduced Coulomb screening, a disordered metal loses the Fermi liquid regime at sufficiently low temperatures. Instead, it presents a DOS vanishing at the Fermi level with $N(E) \propto |E|$ at very small energy scales\,\cite{BK0,BK}. This suggests that the onset of the reduction in Coulomb screening might occur relatively far from the transition and influence the behavior of disordered superconductors. Here we make a comparative study of the low energy normal and superconducting DOS in disordered metallic TiN. We find that the DOS in the normal phase has a strong V-shape anomaly which follows closely the prediction of Refs.\,\cite{BK0, BK} and that the spatial fluctuations of the DOS in the superconducting state occur on the same length scale as those for the  DOS in the normal state.

\begin{figure}[h!]
\begin{center}
\includegraphics[width=0.97\columnwidth]{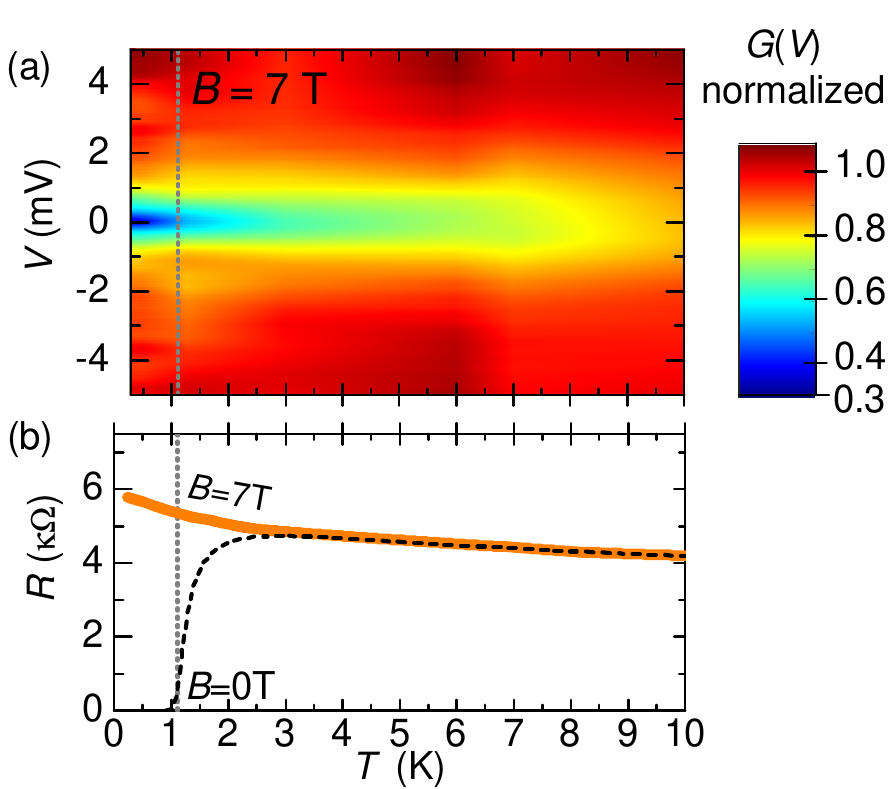}
\caption{(a) The tunneling conductance vs. voltage normalized to its value at 5 mV is shown as a color scale (scale bar on the right) and is plotted as a function of temperature at a magnetic field of 7 T. (b) Sheet resistance vs temperature in the same sample at the same magnetic field (orange line). The result at zero magnetic field is shown by the black dashed line. Vertical dashed line provides T$_c$ at zero magnetic field. The magnetic field is applied perpendicular to the film, well above the critical field. Inset in (b) shows the resistance vs 1/T in a log scale. All resistance vs temperature measurements are provided in the Supplementary Material (Fig.\,S2), the sample is named D15. Well above H$_{c2}$ (of about 2 T at zero temperature), i.e. above about 4 T, the resistance vs temperature and the tunneling conductance is magnetic field independent.} \label{fig1}
\end{center}

\end{figure}

\begin{figure}[th!]
\begin{center}

\includegraphics[width=0.5\textwidth]{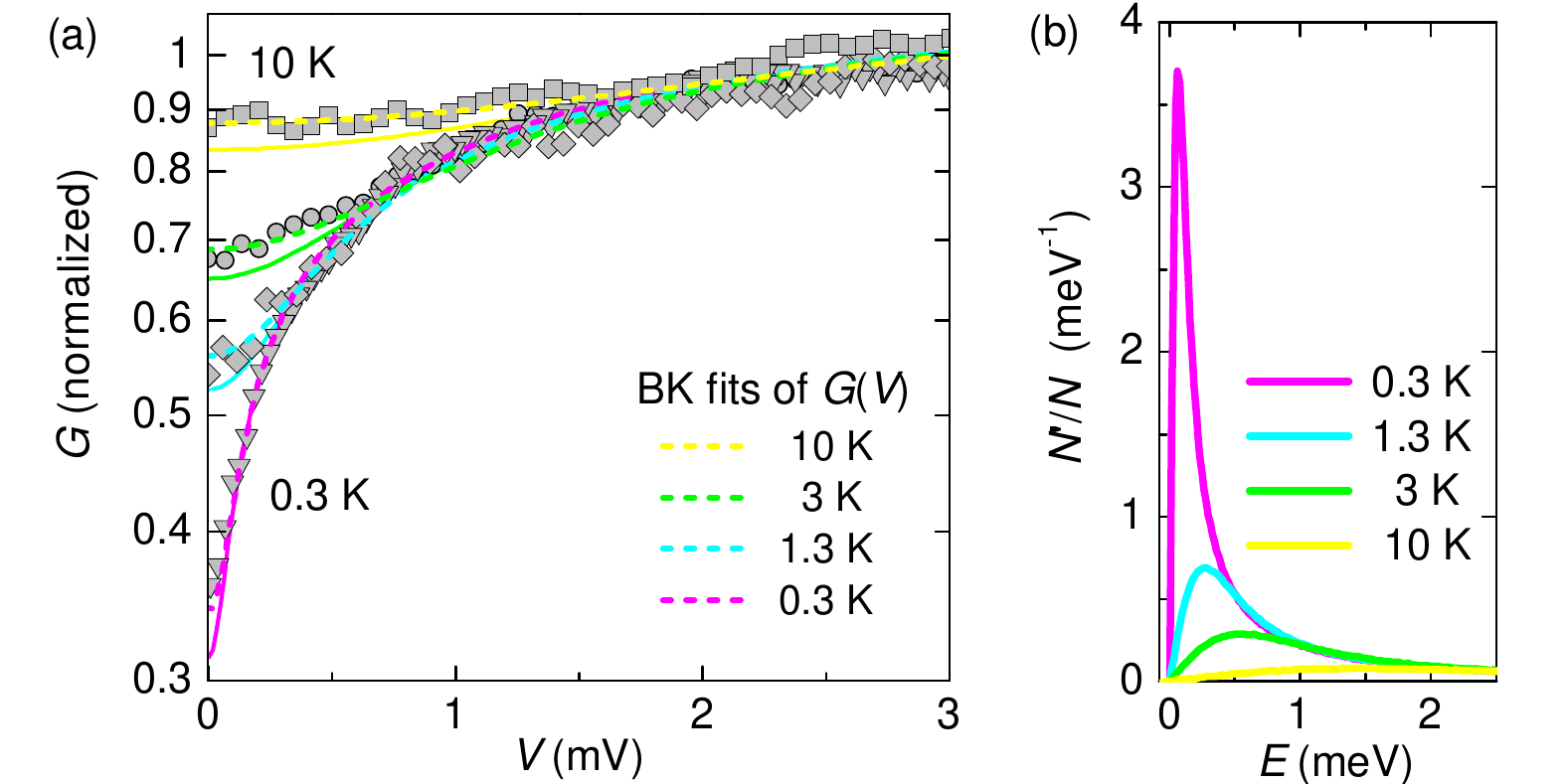}	

\caption{In (a) we show the tunneling conductance $G(V)$ as grey filled symbols for sample D15 (a) normalized to about 3 mV for different temperatures, marked in the figure and a magnetic field of 4 T. Dashed lines provide the fits of the tunneling conductance discussed in the text. Lines show the corresponding DOS, $N(eV)$, obtained after eliminating the temperature induced smearing in $G(V)$. In (b) we show the derivative of the DOS $\frac{N'}{N}=\frac{dN(E)/dE}{N(E)}$ as a function of temperature. In the Supplementary Material we provide the same data for another more resistive sample (Fig.\,S3,S4).
} \label{fig2}
\end{center}
\vspace{-0.9cm}
\end{figure}

We study a 5 nm thick TiN film deposited on a Si/SiO$_2$ substrate at $T$ = 350\,$^{\circ}$C by atomic layer deposition, whose transport properties were chosen to be as close as possible to film TiN2 and TiN3 in Ref.\,\cite{TiNInhom}. A detailed analysis of the morphology of the film using X-ray, Transmittion Electron Microscopy (TEM) and atomic scale Scanning Tunneling Microscopy (STM) is provided in the Supplementary Material (Fig.\,S1). The film consists of a network of strongly interconnected single crystal like structures oriented randomly on the substrate and separated by only small angle grain boundaries. The average distance between boundaries is of 5 nm. The crystal lattice parameters are the same as bulk TiN. The scanning tunneling  microscopy (STM) and spectroscopy results were carried out using a home made dilution refrigerator STM with a voltage resolution of about 15\,$\mu$V\,\cite{Rodrigo04,Guillamon07,Suderow11}. The junction resistance was always well above a M$\Omega$, much larger than the film's sheet resistance. We term our sample D15. We also provide in the Supplementary Material results on the same sample, but with an additional heat treatment (2 minutes heating to 200$^{\circ}C$ on air, Figs\,S3,S4). The heat treatment lead to more insulating behavior. In D15, we performed atomic scale STM and spectroscopic studies at a magnetic field of 4 T and at zero field. We studied a grid similar to the one in Ref.\,\cite{TiNInhom}, with about one tunneling conductance curve each 5 nm, i.e. on average about one curve between small angle crystal boundaries. After the STM measurements, we performed transport experiments, obtaining the results expected from Ref.\,\cite{Baturina12EPL}. For the heated sample, discussed in the Supplementary Material (Figs\,S3,S4), we were not able to scan or obtain atomic resolution, probably due to surface degradation from heat treatment. However, we could obtain tunneling conductance curves at a few locations and the results are consistent with the transport experiments. In the following, we focus mainly on results in D15.

Figure\,\ref{fig1} shows  the  temperature evolution of tunneling conductance $G(V,T)$ together with the temperature dependence of resistance $R(T)$ (solid line in Fig.~\ref{fig1} (b)) under magnetic fields in the normal phase. Here, T$_c=1.1$ K and $H_{c2}=2.6$ T. The tunneling conductance  $G(V)$ (Fig.~\ref{fig1}\,(a)) shows a very strong suppression near zero bias which evolves when decreasing temperature. The suppression goes in parallel to the increase of the resistance $R$ when decreasing temperature. The noticeable suppression of $G(V)$ near $E_F$ persist up to 10 K, as well as the negative slope in $R(T)$.

In figure~\ref{fig2}\,(a) we show the tunneling conductance $G(V)$ for different temperatures. Note that there is a strong decrease of the tunneling conductance towards zero bias. To analyze this behavior in more detail we remind that the sample can be considered as a 2D metal in this temperature range~\cite{Valles89,Baturina12EPL}. There is indeed a crossover from 3D to 2D behaviour when the diffusion length, $2 \pi \hbar D/ E$, approaches the film thickness~\cite{Valles89}. Taking the film thickness $d$ and diffusion coefficient $D$ (see Table in Supplementary Material), we obtain $E \lesssim 5$meV, which is the voltage range we work with. We use the model of Bartosch and Kopietz (BK)~\cite{BK0,BK} which, as we will show, provides an accurate description of our data. The model treats a low-dimensional metal with long-range Coulomb interactions in the limit of small $eV$. The BK model was used for various 1D systems, in which a dip around zero-bias was observed in the tunneling conductance~\cite{1dBK1, 1dBK2, 1dBK3}. The model starts describing a usual metal and takes into account the absence of Coulomb screening when correlations are enhanced by disorder. The model provides the DOS as a function of temperature, showing a dip which develops when decreasing temperature. The energy scale for the dip is related to the scattering parameters, particularly $\tau_0$, the defect scattering time and an additional time related to electron-electron interactions, $\tau_1$. We can write, following~\cite{BK0,BK}, that the tunneling conductance is given by
\begin{eqnarray}
G(V, T) \propto \coth \left(\frac{V}{2T}\right)2T
\int^{\infty}_{\tau_0}dt\frac{\sin(V t)}{\sinh(\pi t T)} \, \, \,  \,  \, \,  \, \, \, \nonumber \\
\, \,  \,  \,  \,  \, \, \times \exp
\left[-\frac{r_0}{4}\ln\left(\frac{t}{\tau_1}\right)\ln\left(\frac{t}{\tau_0}\right)\right],
\label{DOS}
\end{eqnarray}
%
where we have taken into account that the $G(V)$ is the DOS smeared by the Fermi function. $r_0$ is a dimensionless measure for system's  resistance~\cite{BK} and $\tau_1 = 4 \tau_0/(\kappa l)^4$ where $\kappa$ is the Thomas-Fermi screening wave vector. For a good metal, the Thomas-Fermi screening length is short compared with the mean free path ($\kappa l \gg 1 $) so that $\tau_1 \ll \tau_0$ and we find flat $G(V)$ and an energy independent DOS~\cite{BK}. In our case, $\tau_1$ and $\tau_0$ are of the same order, leading to the observed ZBA~\cite{BK}.

In Fig.\,~\ref{fig2}(a) we show as dashed lines the temperature evolution of  $G(V)$ together with curves calculated with Eqs.~(\ref{DOS}). Notice the similarity to the actual DOS, shown by the lines in Fig.\,~\ref{fig2}(a), indicating that temperature smearing is weak as compared to the effect of Coulomb correlations. We use just two parameters, $r_0$ and $\tau_0$ (taking $\tau_1=\tau_0$), which already provides an excellent fit. We find $\tau_0= 7.3 \pm 0.3 \cdot 10^{-15}$\,s. We this, we find $R=3 \pm 0.05$\,k$\Omega$ (from $r_0 = R/(h/e^2)$), which is very close to sample's sheet resistance at room temperature $R_{\Box} = 2.94$\,k$\Omega$. Furthermore, this value also $\tau_0$ describes the suppression of the superconducting critical temperature $T_c$, following an expression originally proposed by Finkel'stein~\cite{Fink, Postolova2017} (see also Supplementary Material Fig.\,S5). The suppression of $T_c$ is due to scattering and electron density fluctuations produced by the reduction of Coulomb screening. As we will see below, the length scale associated to spatial DOS fluctuations we measure here is related to $\tau_0$.

BK model exemplarily shows how the Fermi liquid regime is lost by Coulomb interactions in low dimensionality. In essence, the Coulomb gap  $N (E) \propto |E|$ appears in the DOS in a system with finite conductance at a finite temperature at energies sufficiently close to the Fermi level~\cite{BK0}.

\begin{figure}[h]
\begin{center}
\includegraphics[width=0.45\textwidth]{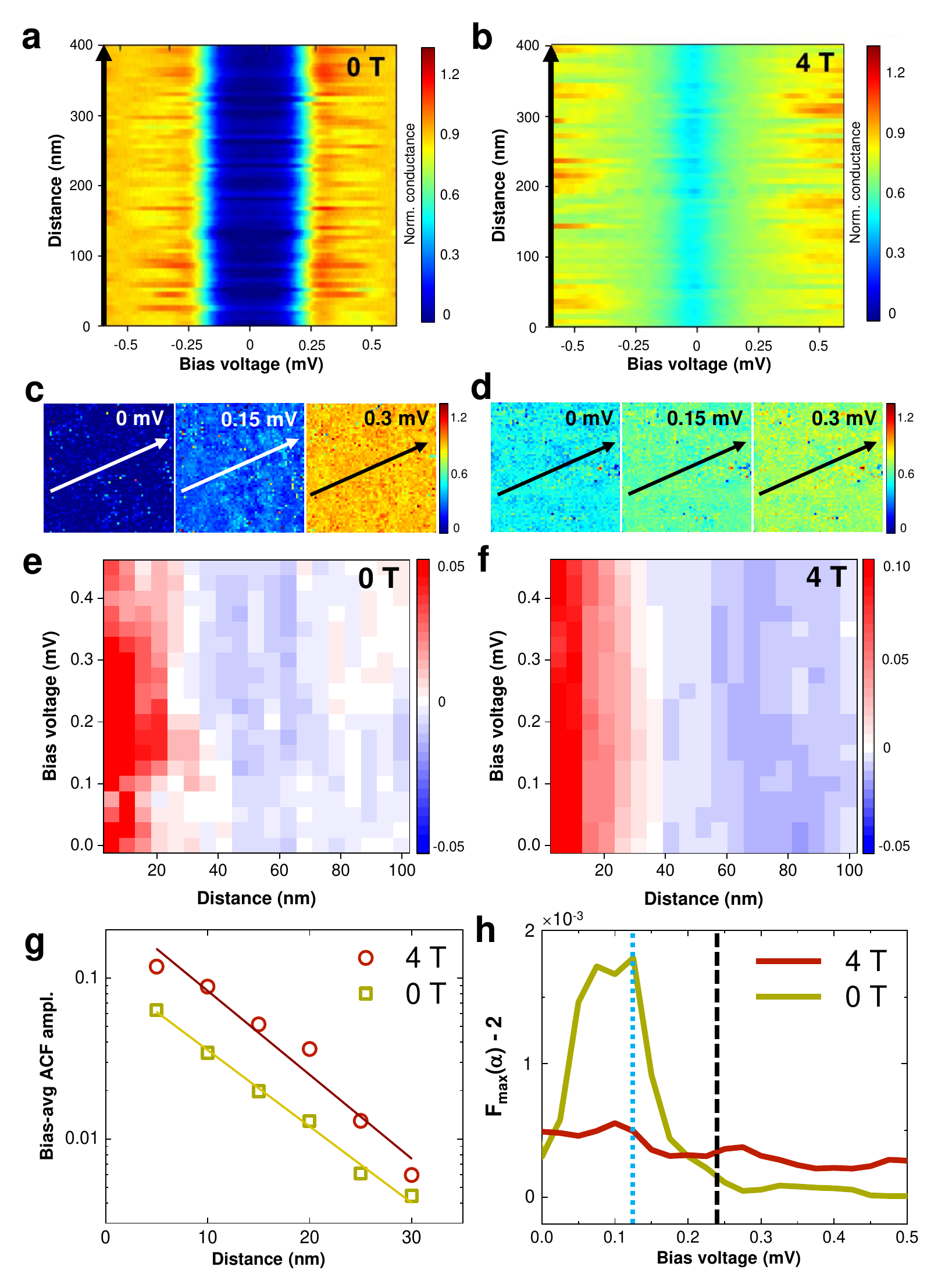}
\vskip -0.2cm	
\caption{In (a,b) we show the tunneling conductance normalized to its value at 1 mV as a function of the position along a line (black arrow) at zero field (a) and with a magnetic field of 4 T (b). The color scale provides the normalized conductance following the scale bar in the right of each panel. In (c,d) we show tunneling conductance maps over an area of 400 nm at zero magnetic field (c) and with a magnetic field of 4 T (d). The field of view is slightly different, due to a modification of the position when applying the magnetic field. We show maps at the 0 mV, 0.15 mV and 0.3 mV, from left to right. The normalized conductance is shown by the bars in the right side of each figure. The arrows provide the scan over which we trace the tunneling conductance in (a,b). Note that we make one tunneling conductance curve approximately each 5 nm. In (e,f) we show the $ACF(r)$ at zero field and at 4 T. The amplitude of the ACF is given by the color bar on the right of each figure. In (g) we show $ACF(r)$ averaged for all bias voltages at zero magnetic field (brown squares) and at 4 T (red circles). In (h) we show the position of the maximum of the distribution function of multifractal exponents $F(\alpha)$ as a function of the bias voltage. We mark by a black vertical dashed line the position of the quasiparticle peaks and by a blue dashed line the position of the maximum in $F(\alpha)$.} \label{fig3}
\end{center}

\end{figure}

Notice that we are in the peculiar situation where the normal state DOS $N(E)$ varies in the same energy scale as the superconducting gap. The shift in the chemical potential occuring in BCS theory below $T_c$ ($\mu_0-\frac{\Delta^2}{\mu_0}$ where $\mu_0$ is the chemical potential of the normal phase) is then modified by a factor $\frac{N'}{N} (E_F)$, giving $\mu_0-c\frac{N'}{N}(E_F) \Delta^2$ (with $c$ being a constant)\,\cite{dRo}. $E_F$ is taken in this simple approach to be the Fermi level. In Fig.\,~\ref{fig2}(b) we show $\frac{N'}{N} (E)$ for different temperatures. We see that $\frac{N'}{N} (E)$ is a nonmonotonic function with a well developed maximum that rapidly increases with cooling. The energy where the maximum occurs is decreasing with temperature. For temperatures below $T_c$, the maximum is very close to the Fermi level. We can take the point where $\frac{N'}{N} (E\approx E_F)$ is not zero as determining the shift in the chemical potential. This provides in our case a value that is considerably larger than the one found for noninteracting electrons. Furthermore, it increases most when crossing T$_c$ towards lower temperatures. We should note that the value of the superconducting gap is also considerably larger than the BCS value $\approx 1.76 k_BT_c$\cite{TiNInhom}. This suggests that there could be a link between the increased shift in the chemical potential due to the shape of the normal DOS and the superconducting gap value. On the other hand, at temperatures well above $T_c$, the maximum in $\frac{N'}{N}(E)$ is shallow and separated from the Fermi level. Thus, the reduction of Coulomb screening and the decrease in the Fermi level DOS prevents the formation of Cooper pairs, explaining the decrease in $T_c$ when increasing disorder. The same tendency is obtained for sample D15e, as shown in the Supplementary Material (Figs.\,S3,S4).

The normal phase DOS, according to BK, might also show small spatial fluctuations, if the scattering is spatially dependent, which implies a spatially dependent $r_0$\cite{BK0,BK}. The morphology of the thin films, which consist of slightly misoriented crystals connected through small angle boundaries (Supplementary Material Fig.\,S1), suggests indeed that scattering can be strongly spatially dependent. Here we find fluctuations in the normal phase tunneling conductance. These connect to the zero field DOS fluctuations.

In Fig.\,~\ref{fig3}(a-d) we show the spatially fluctuating tunneling conductance acquired in the normal and superconducting phases. The fluctuations in the tunneling conductance amount to small variations of a few \% as a function of the position. To obtain a length scale associated to these fluctuations we calculate the autocorrelation function of the tunneling conductance images. The autocorrelation function $ACF (\vec{r})$ provides the statistical correlation between any two points in an image separated by a vector $\vec{r}$ (see Refs.\,\cite{Fratini2010,Giraldo-Gallo2015} and Supplementary Material, Fig.\,S6). As there are no symmetric patterns in our images, we plot the radially averaged $ACF(r)$ in Fig.\,~\ref{fig3}(e,f). We observe that $ACF(r)$ decreases exponentially with $r$. When we average over all bias voltages, we find $ACF(r)\propto e^{-\frac{d}{r}}$ with $d\approx$ 20 nm (Fig.\,\ref{fig3}(g)). Thus, we identify a length scale $d\approx$ 20 nm, which is shared by normal and superconducting phases. Taking a value of $v_F=2\times 10^6 m/s$ consistent with bandstructure calculations for bulk TiN\,\cite{Neckel_1975} and the above value of $\tau_0= 7.3 \pm 0.3 \cdot 10^{-15}$\,s, we find that $d \approx v_F \tau_0$. Thus, the depression of the DOS induced by Coulomb interactions also leads to a spatial fluctuations in the DOS.

The detailed spatial distribution of the conductance suffers however slight changes when entering the superconducting phase. The absolute value of $ACF(r)$ is slightly larger at 4 T than at 0 T, indicating that the superconducting phase leads to a modification of the correlations. There is a decrease in the size of $ACF$ inside the superconducting gap, although the length scale for the decay of $ACF(r)$ remains of the same size. This shows that the variations of the conductance are stronger in the superconducting phase. To characterize the spatial distribution of these variations, we have calculated the distribution function of multifractal exponents $F(\alpha)$ (we follow the standard procedure to obtain multifractal exponents, described in detail recently in Appendix F of Ref.\,\cite{Benito19}, the code can be obtained at \cite{codeFractal}). In presence of random variations, $F(\alpha)$ is only narrowly defined around $\alpha=2$. If the spatial variations show some structure, $F(\alpha)$ broadens and becomes an inverted parabola whose maximum value deviates from $2$. This serves to identify a tendency to form droplets in the image that are barely visible directly. This tendency shows that the DOS distribution is random in the normal phase but shows some spatial structure in the superconducting phase. This occurs particularly for bias voltages of about half the quasiparticle peak position (Fig.\,\ref{fig3}(h)) and suggests that there is an emergent granularity which becomes more pronounced inside the superconducting phase. We can speculate that the tendency continues when approaching and crossing the SIT, with a shift in the granularity towards the gap edge\cite{Feigelman07,FEIGELMAN20101390}.

Recently, authors of Ref.\,\cite{carbillet2019spectroscopic} analyzed the DOS for energies above the superconducting gap and found a connection between the Coulomb-like variation of the normal DOS and the size of the superconducting gap in thin NbN films. Since the upper critical field was too high, the low energy DOS could not be established. Nevertheless, they found a neat anticorrelation between the power law of the ZBA and the size of the superconducting gap, which is not related to the topography of the sample. With this, they showed that the Coulomb interactions modify the value of the superconducting gap and are responsible for the granularity in the superconducting phase. Here we find instead that there are random variations of the ZBA and which remain overall random in the superconducting phase, although with a tendency to form spatial structures (Fig.\,\ref{fig3}(h)).

To sum up, we observe a severe enhancement of the suppression of DOS near Fermi energy with decreasing temperature in the normal phase of a quasi-2D disordered thin film. We show a quantitative agreement between experiment and theory developed for a two-dimensional metal with long-range Coulomb interactions~\cite{BK0, BK}. Our results experimentally verify the calculations showing that the Fermi liquid regime is lost in presence of Coulomb correlations. The sharp $N (E)$ strongly modifies the establishment of superconducting correlations. It is spatially fluctuating, with a length scale shared by normal and superconducting phases and there is an incipient formation of structure inside the superconducting phase.

\section{Acknowledgments}
The work at Universidad Aut\'onoma de Madrid was supported by
the Consejer\'{i}a de Educaci\'{o}n, Cultura y Deporte (Comunidad de Madrid)
through the talent attraction program.
The work at Argonne (V.M.V.) was supported by the U.S. Department of Energy, Office of Science, Basic Energy Sciences, Materials Sciences and Engineering Division.
The work of T.I.B. on the experiment was supported RSF project No. 19-72-30023.
The work of S.V.P. on the analysis of experimental data was supported by RFBR project No. 18-32-00718 mol-a.
We acknowledge discussions with P. Kopietz and Ch. Strunk. We particularly acknowledge intense discussions with S. Vieira. We also acknowledge help for data acquisition by P. Kulkarni and for calculations by G. Martinez. We acknowledge support by the Spanish Research State Agency (FIS2017-84330-R and “Mar\'ia de Maeztu” Programme for Units of Excellence in R\&D CEX2018-000805-M), by the Comunidad de Madrid through program NANOMAGCOST-CM (Grant No. S2018/NMT-4321) and by EU program Cost CA16218 (Nanocohybri). We also acknowledge the SEGAINVEX at UAM.

\newpage

\section{Note A: Structural characterization of TiN thin films}

We made a characterization of the morphology of the film studied in the main text (D15) by combining x-ray, TEM and STM results.

\begin{figure*}[h]
\begin{center}
\includegraphics[width=0.8\textwidth]{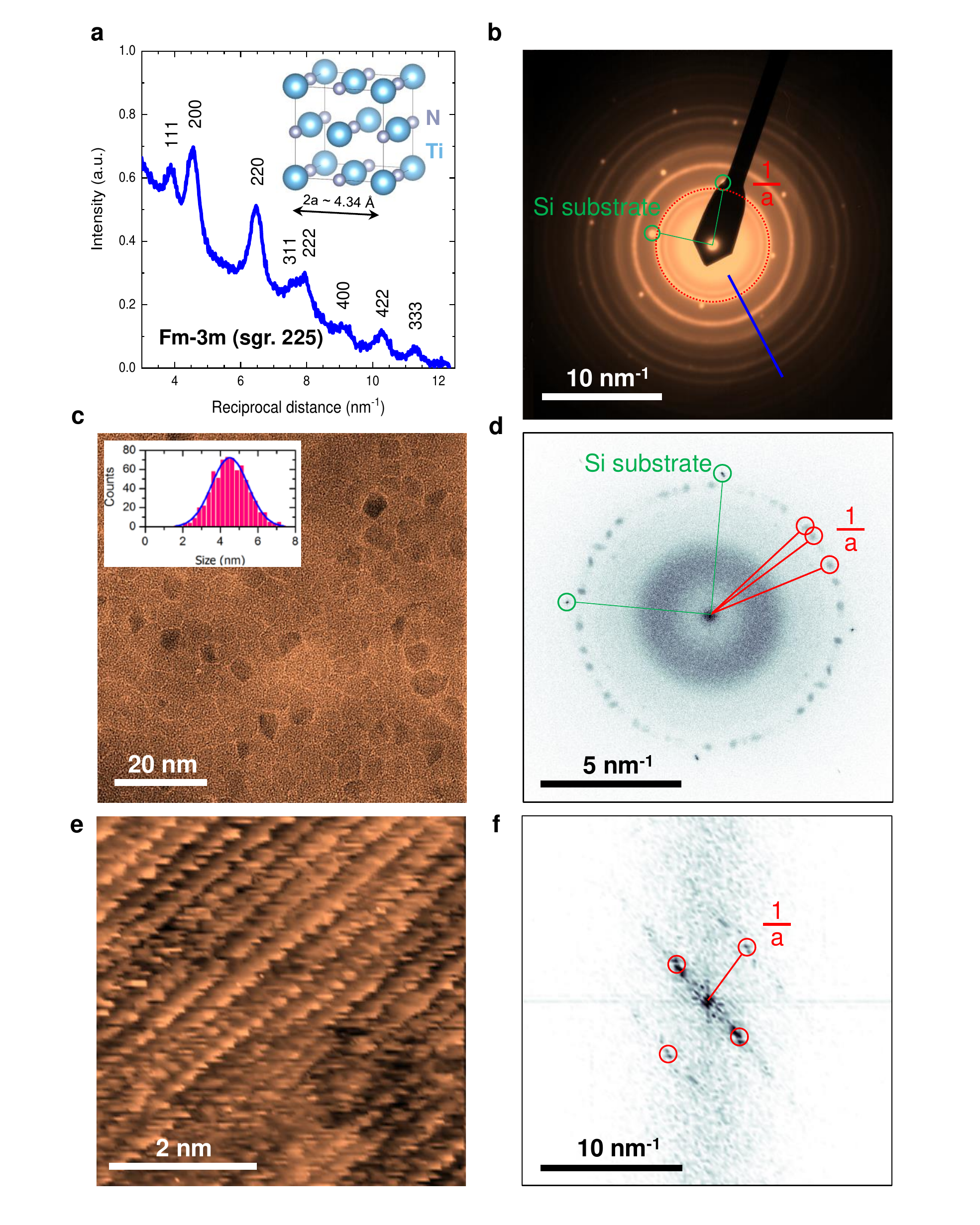}	
\caption{In a we show the intensity vs reciprocal distance obtained in x-ray scattering of a thin film of TiN. We index each peaks with their corresponding lattice planes. The space group is given in the legend. The structure is depicted in the upper right part of the figure. In (b) we show the actual x-ray scattering experiment. The blue line provides the scan made to obtain (a). The green circles show the single crystalline Si substrate. We mark with a red dashed line circle the reciprocal size of the lattice constant $a$, as defined in the upper right inset of (a). In (c) we show a Transmission Electron Microscopy (TEM) image of the TiN film. We observe that the film consists of a TiN crystals separated by small angle grain boundaries. We measured the distance between boundaries and show the result as a histogram in the inset. In (d) we show the Fourier transform of (c). We remark the peaks corresponding to the Si substrate by green circles. The spots coming from the crystallites are distributed on a circle. We mark the reciprocal lattice constant $a$ by red circles. In (e) we show a STM image of a single grain of the TiN film. We observe the atomic lattice. The corresponding Fourier transform is shown in (f). We highlight the reciprocal distance $a$ again with red circles and a line.} \label{figStructure}
\end{center}
\vspace{-0.9cm}
\end{figure*}

If there is crystalline order at some length scale, X-ray scattering leads to rings located at Bragg peaks of the lattice constant. Indeed, X-ray scattering of our thin films leads to ring like patterns, which are superposed to the four-fold crystalline Bragg peaks of the Si substrate (Fig.\,\ref{figStructure}(a,b)). We can index all peaks with the planes corresponding to the bulk crystal structure (Fig.\,\ref{figStructure}(a)). This shows that growth favors the nucleation of the bulk crystalline structure.

When analyzing the films with TEM, we observe a distribution of many small angle boundaries between small crystals of slightly differing orientations (Fig.\,\ref{figStructure}(c)). The corresponding Fourier transform (Fig.\,\ref{figStructure}(d)) shows again the Bragg peaks of the substrate and those of the crystalline structure of TiN. Because the field of view is limited, we observe many Bragg peaks, all located at the same distance, instead of the ring that is observed for the whole film in x-ray scattering. Notice the small difference between orientations of crystals. Analyzing the TEM image, we can obtain the distribution of the distance between boundaries. As we show in the inset of (Fig.\,\ref{figStructure}(c)), we find a broad distribution, starting at about 2 nm and ending at about 7 nm. The average distance between small angle boundaries is of 5 nm.

In a STM experiment, we can zoom into a small single crystal, as shown in Fig.\,\ref{figStructure}(e). We can observe the atomic lattice. The square lattice is somewhat distorted due to disorder or tip related effects during scanning, but in the Fourier transform, Fig.\,\ref{figStructure}(f), we identify again peaks at the interatomic distances.

Thus, in all, we have a highly disordered system. Scattering centers are unlikely coinciding always with small angle boundaries, so that the relevant length scale can be much larger, as we actually show from our experiments and discuss in the main text.

\section{Note B: Transport in the sample of the main text and in another, more insulating, sample}

\begin{figure*}[tbh]
    \begin{center}
        \includegraphics[width=0.8\linewidth]{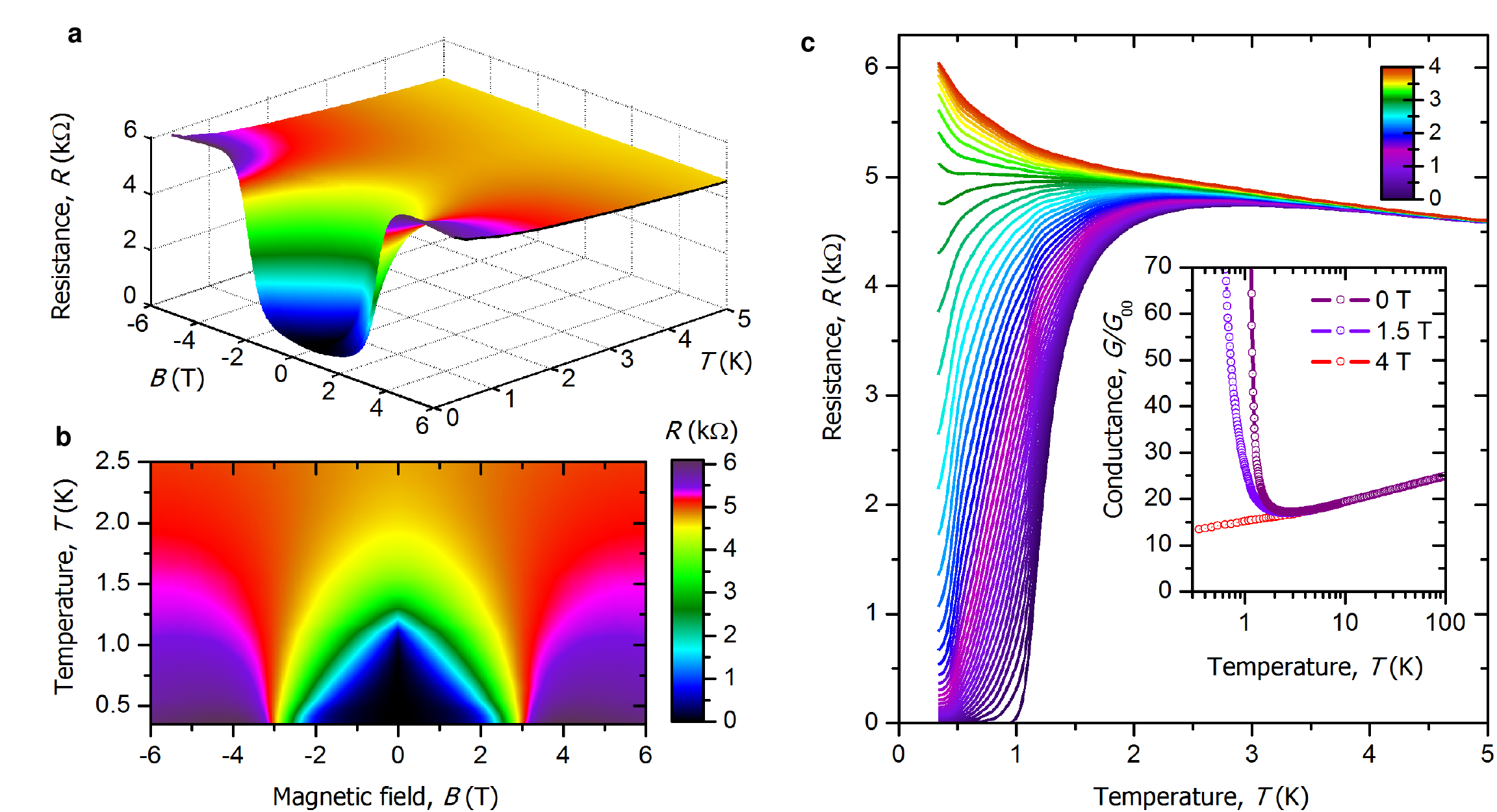}
    \end{center} \vspace{-5mm}
\caption{
        In (a) we show the temperature and magnetic field dependence of the sheet resistance of the TiN film discussed in the main text. In (b) we show the same data with the resistance plotted in a color scale. The bar providing the scale for the resistance is given in the right of (b). In (c) we show the resistance vs temperature for different magnetic fields. The color corresponds to the value of the magnetic field, following the bar on the upper right. In the inset we show the conductance as a function of temperature (normalized to the quantum of conductance $G_{00}$), in a logarithmic scale.}\label{ResistanceD15}
\end{figure*}

It is useful to show and discuss the whole temperature and magnetic field dependence of the resistance. In Fig.\,\ref{ResistanceD15} we show the result for the sample discussed in the main text. Notice that the upper critical field is of approximately 2.6 T and that the variation of the resistance with the magnetic field strongly flattens out for fields above the critical field. The resistance (inset of Fig.\,\ref{ResistanceD15}(c)) shows an increase with decreasing temperature in the normal phase above 4 T.

\begin{figure*}[tbh]
	\begin{center}
		\begin{center}
			\includegraphics[width=0.87\linewidth]{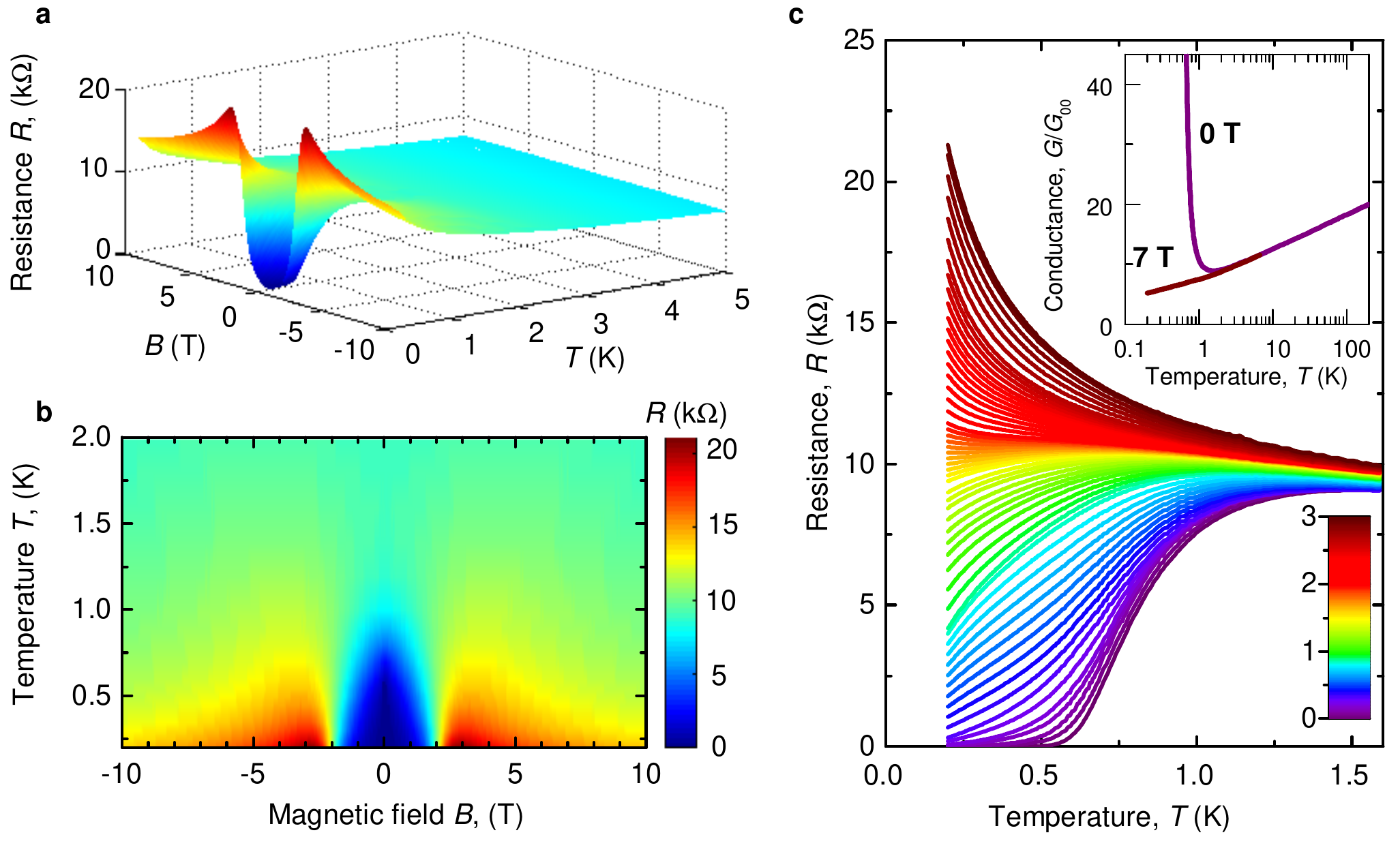}
		\end{center}
		\vspace{-0.3cm}
		\caption
		{In (a) we show the sheet resistance vs temperature and magnetic field of a thin film with a larger normal phase resistance than the one discussed in the main text. In (b) we show the corresponding colormap, with the resistance given as a color. The scale follows the bar on the right. In (c) we show the resistance as a function of temperature for different magnetic fields, with the magnetic field given by the bar on the bottom right. The top right inset is the normalized conductance $G/G_{00}$ vs. temperature at zero magnetic field and at $B=7$\,T.}\label{ResistanceD15e}
	\end{center}
\end{figure*}

In  Fig.\,\ref{ResistanceD15e} we show the results other film mentioned in the text, D15e, which was obtained from the same film after an additional heat treatment. We observe that the critical temperature and the upper critical field are considerably smaller. At the same time, the resistance shows a stronger increase with decreasing temperature in the normal phase under magnetic fields.

\begin{figure*}[tbh]
    \begin{center}
        \includegraphics[width=0.99\linewidth]{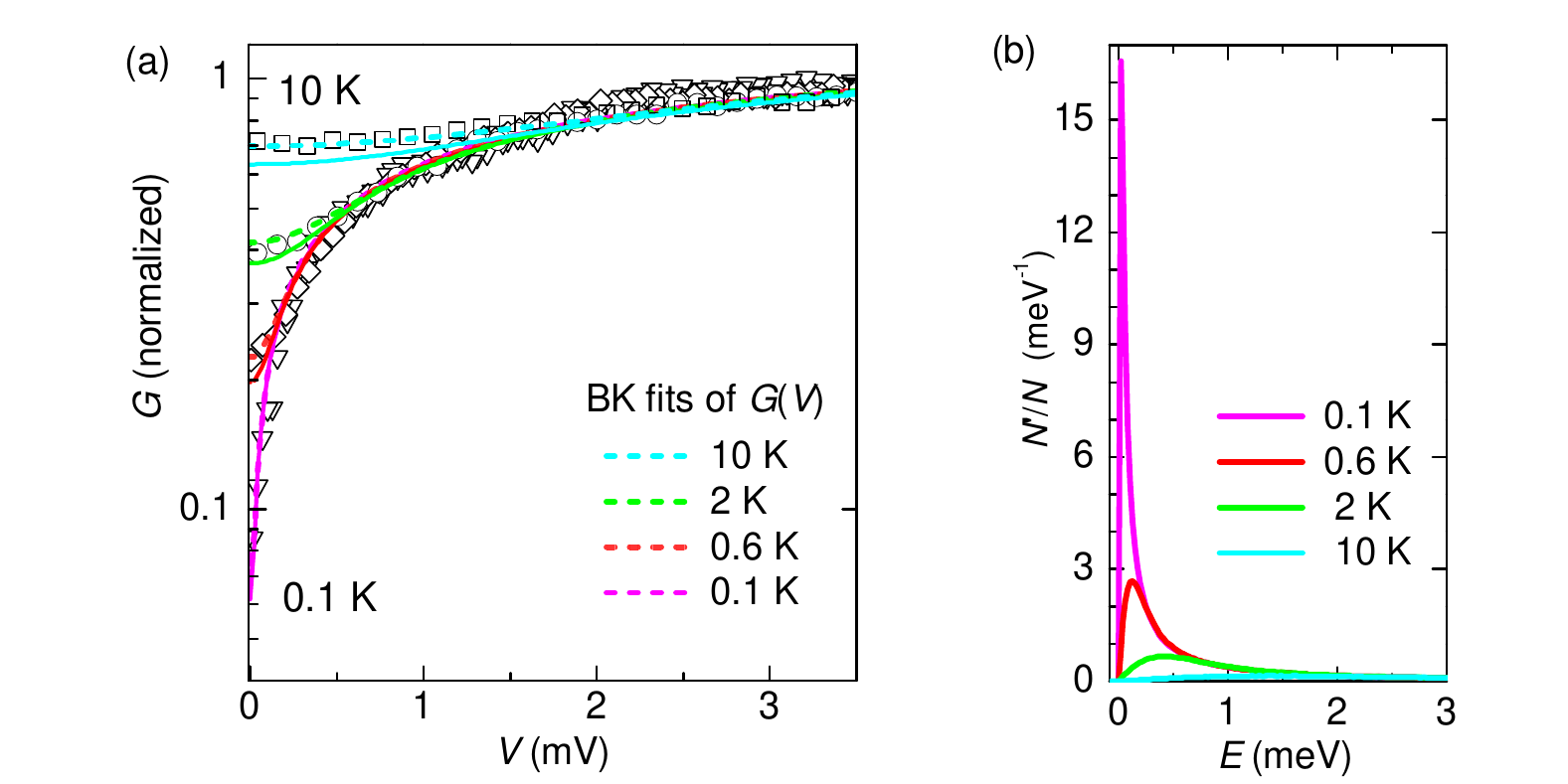}
    \end{center} \vspace{-5mm}
\caption{In (a) we show the temperature evolution of the tunneling conductance $G(V)$ for a sample which is more resistive than the one discussed in the main text. The sample is termed D15e. We show the data as black circles. The fits to the tunneling conductance $G(V)$ are given as dashed lines. The DOS according to Bartosh-Kopietz is given by the lines. In (b) we show the derivative of the DOS for different temperatures.}\label{TunnelingD15e}
\end{figure*}

In the sample D15e we took several tunneling conductance curves. We show the result in Fig.\,\ref{TunnelingD15e}. We only show data in the normal phase obtained by applying a magnetic field of 7 T. The superconducting gap opens as a tiny feature in the center of the curve and we could not analyze this in detail. We see that in this sample the tunneling conductance drops to zero at low temperatures much more strongly than in D15. The behavior is also captured by BK model, which provides a temperature variation of $\frac{N'}{N}$ 
(Fig.\,\ref{TunnelingD15e}(b)) leading to a peak at energies that are slightly smaller, but of the same order as in D15.

We summarize some parameters obtained previously in both samples in Table\,S\ref{TableParameters}.

\begin{figure}[tbh]
    \begin{center}
        \includegraphics[width=0.99\linewidth]{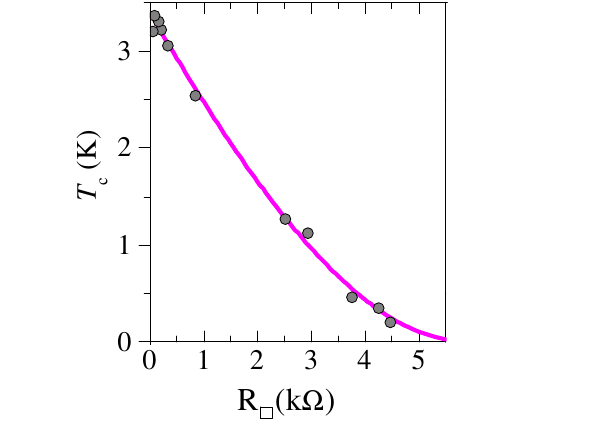}
    \end{center} \vspace{-7mm}
\caption{
        We reproduce data and model of Ref.\,\cite{Postolova2017}. We show the critical temperature vs the sheet resistance in different TiN samples as black points. The magenta line is the same quantity calculated using Ref.\,~\cite{Finkelstein}.}\label{FigFink}
\end{figure}

In the Fig.\,\ref{FigFink} we provide the critical temperature vs sheet resistance in TiN films, from \cite{Postolova2017}. There, we calculated the dependence of $T_c$ using $
\ln \left( \frac{T_{c}}%
{T_{c0}}\right) = \gamma + \frac{1}%
{\sqrt{2r}} \ln \left( \frac{1/\gamma + r/4 - \sqrt{r/2}}%
{1/\gamma + r/4 + \sqrt{r/2}}\right),
$
where $r = G_{00}\cdot R$, $R$ is the normal resistance per
square, $\gamma = \ln [\hbar/ (kT_{c0} \tau)]$. This expression was derived by Finkel'stein to explain the decrease in T$_c$ in disordered thin films due to scattering and electron density fluctuations (Ref.\,24 of the main text). We see that this expression explains the variation of $T_c$ with $R$. By taking $\gamma = 5.73$ we found $ \tau = 7.3\cdot10^{-15}$\,s, the value we obtain in the main text from our fits to the conductance and from the measured spatial dependence of the DOS fluctuations in the normal phase.

\begin{table*}[tbh]
\caption{The sample D15e was made from a sample identical to D15 by heating.
$R_{\Box}$ -- the resistance of the film per square
at room temperature;
$B_{c2}(0)$ ---  the upper critical field at $T=0$;
$D$ -- diffusion coefficient ($D=\left(\frac{\pi}{2\gamma}\right)\left(\frac{k_BT_c}{eB_{c2(0)}}\right)$, with $\gamma=1.781$ being Euler's constant).}
\begin{center}
\begin{tabular}{|c|c|c|c|c|c|}
\hline
Sample & $R_{\Box}$& $B_{c2}(0)$ & $T_c$ &   $D$  \\
         &k$\Omega$  &    T        & K     &cm$^{2}/$s  \\
\hline
D15    & 2.94      &  2.65        &1.12  &   0.32  \\
\hline
D15e   & 3.87      &  2           &0.66  &   0.25    \\
\hline
\end{tabular}
\end{center}\label{TableParameters}
\end{table*}

\section{Note C: Calculation of the autocorrelation function}

In absence of visible patterns, the most efficient way to define a length scale is to calculate the autocorrelation function $ACF$. This was made in Refs.\,\cite{Fratini2010,Giraldo-Gallo2015} to obtain spatial dependencies related to the opening of the pseudogap in cuprates and to analyze patterns in an oxyde sample.

We should distinguish between the autocorrelation function often used to analyze pictures or images and the statistical $ACF$ we use here.

The usual method starts by defining two arbitrary matrices $A$ and $B$\,\cite{matlab}. The cross-correlation matrix $C$ is a measure of similarity between $A$ and $B$ as a function of the displacement of one relative to the other. The elements of $C$ are calculated by displacing A over B a given vectorial lag, then calculating the element to element product of the overlapping elements and finally taking the sum of them. If we take a  $2 \times 2$ matrix $A=\begin{bmatrix}a_{11} &a_{12}\\a_{21}&a_{22} \\\end{bmatrix}$ and a $3 \times 3$ matrix $B=\begin{bmatrix}b_{11} &b_{12} &b_{13}\\b_{21} &b_{22} &b_{23}\\b_{31} &b_{32} &b_{33}\\\end{bmatrix} $.

\begin{center}
$C=\begin{bmatrix}c_{-1,-1} &c_{-1,0} &c_{-1,1} &c_{-1,2}  \\c_{0,-1} &c_{0,0} &c_{0,1} &c_{0,2} \\c_{1,-1} &c_{1,0} &c_{1,1}&c_{1,2} \\c_{2,-1} &c_{2,0} &c_{2,1} &c_{2,2}  \\\end{bmatrix}$
\end{center}

Each element of $C$ is calculated by displacing $A$ over $B$ according to the corresponding index, being the $(0,0)$ case when both have the $[11]$ element aligned, then multiplying the overlapping region element-by-element and taking their sum. According to this, let's see how to calculate some elements of $C$:

\begin{center}
$c_{0,0}=a_{11}b_{11}+a_{12}b_{12}+a_{21}b_{21}+a_{22}b_{22}$ \\
$c_{2,1}=a_{11}b_{23}+a_{21}b_{33}$\\
$c_{-1,-1}=a_{22}b_{11}$\\
$c_{2,2}=a_{11}b_{33}$
\end{center}

Thus, the self correlation of a matrix is just the cross-correlation of the matrix with itself.

However, the statistical $ACF$ measures correlations between two points separated by a distance $r$. The spatial autocorrelation function $ACF(r)$ of an image is given by the correlation of any two pixels $i$ and $j$ of the image, separated by a given vector $\bf r = ($r$,\theta)=\bf r_i - \bf r_j$, where $\bf r_i$ and $\bf r_j$ the position of those pixels\,\cite{Fratini2010,Giraldo-Gallo2015}. This leads to the following definitions:

\begin{equation}
ACF({\bf r}) = \frac{1}{N({\bf r})} \sum_{i,j}\frac{(I_i-\langle{I}\rangle_1)(I_j-\langle{I}\rangle_2)}{\sigma_1\sigma_2}
\label{Eq. 1}
\end{equation}
where
\begin{equation}
N({\bf r})=\sum_{i,j}\delta_{\bf r,(r_i-r_j)}
\label{Eq. 2a}
\end{equation}
\begin{equation}
\langle{I}\rangle_1=\frac{1}{N({\bf r})} \sum_{i,j}\delta_{\bf r,(r_i-r_j)}I_i
\label{Eq. 2b}
\end{equation}
\begin{equation}
\langle{I}\rangle_2=\frac{1}{N({\bf r})} \sum_{i,j}\delta_{\bf r,(r_i-r_j)}I_j
\label{Eq. 2c}
\end{equation}
\begin{equation}
\sigma_1^2=\bigg( \frac{1}{N({\bf r})} \sum_{i,j}\delta_{\bf r,(r_i-r_j)}I_i^2 \bigg) -(\langle{I}\rangle_1)^2
\label{Eq. 2d}
\end{equation}
\begin{equation}
\sigma_2^2=\bigg( \frac{1}{N({\bf r})} \sum_{i,j}\delta_{\bf r,(r_i-r_j)}I_j^2 \bigg) -(\langle{I}\rangle_2)^2
\label{Eq. 2e}
\end{equation}

The radial averaged spatial autocorrelation function $ACF(r)$ is the result of averaging the autocorrelation function value for all the vectors with the same magnitude $r=|\bf{r}|$. It gives an idea of the spatial extension of the correlated regions in the image. A white noise image has a flat and close to zero $ACF(r)$, with however a sharp peak at $r=0$, because every pixel is correlated with itself. In real images, peaks $ACF(r)$ mean that there are regions where pixel intensities are spatially correlated, within the same region or with neighbouring regions.

The often used correlation method for images is computationally very fast, since it just needs to "slide, multiply and sum" the matrix over itself a total of $(2d-1)^2$ times, being $d$ the dimension of the matrix. However, it is extremely sensitive to offsets and does not produce the statistical correlation among points exactly. For the calculation of one element it blindly multiplies and sums neighbouring pixels  regardless of their relative intensities, orientation or distance. This can cause that two actually correlated elements or regions vanish or are obscured because of negative-plus-positive sums of their surrounding region. Also, it has no renormalization or consideration for the bright center. When images are only slightly displaced they have much more overlapping pixels contributing to the sum than when they are barely touching, making the border of the self correlation matrix fainter.

The calculation of the statistical $ACF(r)$ \cite{Fratini2010,Giraldo-Gallo2015} by contrast is an offset-independent method, i.e. the result depends only on the relative difference of intensity between two pixels, not on their absolute value. It does not have the drawbacks mentioned in the previous paragraph and provides the connections present between points of a certain distance. The main drawback is of course that it is computationally much more demanding, because it individually checks, counts, multiplies and classifies every possible pair of pixels on the image several times, in order to calculate every parameter needed for the sums.

\begin{figure*}[h]
\begin{center}
\includegraphics[width=0.95\textwidth]{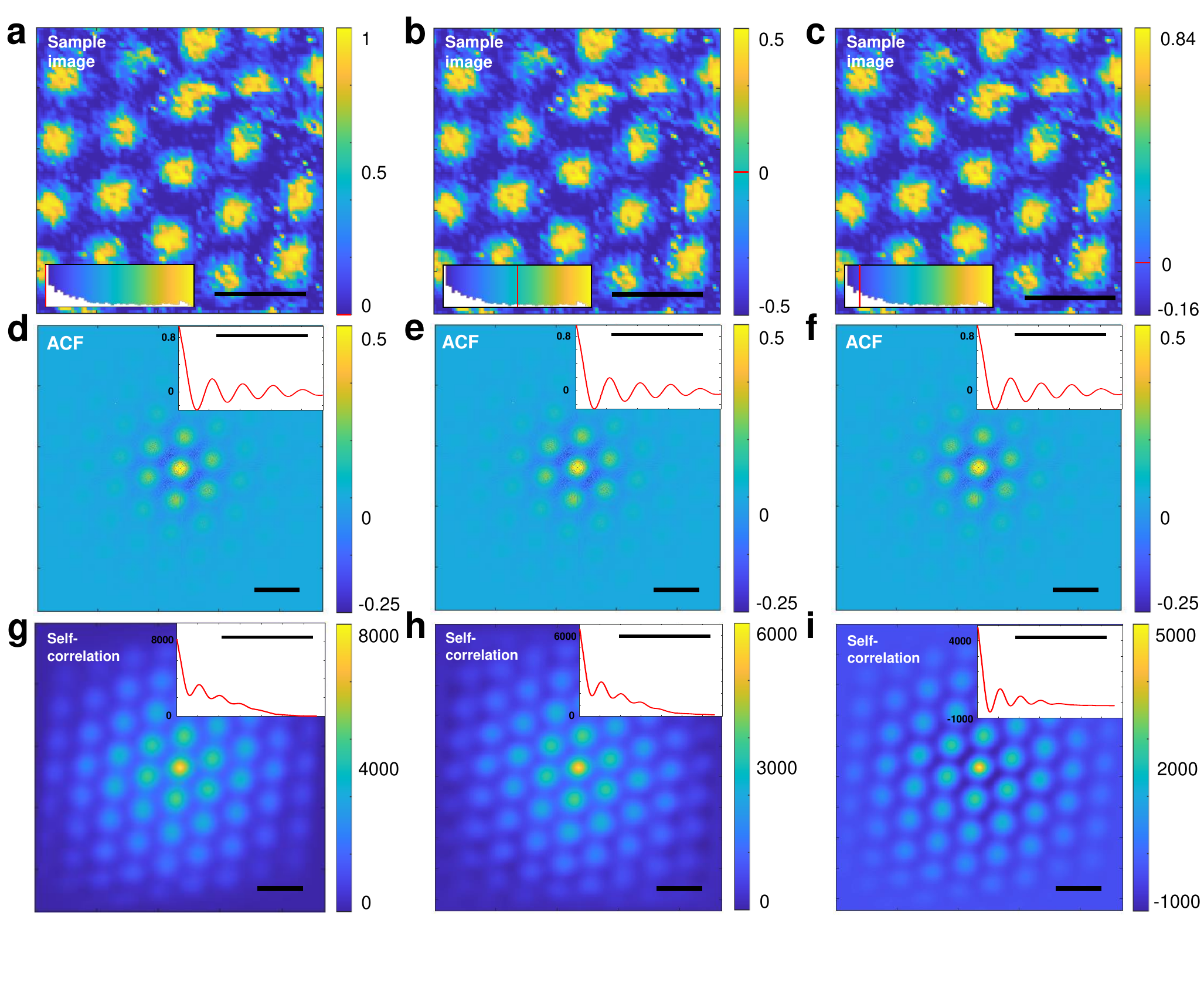}	
\caption{In a-c we show a test image with a vortex lattice, from Ref.\cite{FentePhD}. The color scale is given by the scale bar at the right in arbitrary units. The inset shows the color scale (from blue to yellow) and in white the color histogram from the image. The red line shows the position of zero in the image. In (a), the bottom of the scale is given the zero value, in (b) the center of the scale and in (c) the position of the average pixel value. In (d-f) we show the result of the statistical $ACF(r,\phi)$ following the model described in the text for each of the images in (a-c). The bar in the right of each image provides the amplitude of the $ACF(r)$. In the insets we show $ACF(r)$. We see that all figures (d-f) are identical. In (g-h) we show the correlation function obtained applying usual image analysis software to the images (a-c). Insets provide the radial average and the color bar at the right the amplitude. The result is strongly dependent on the average value of the color scale of the image. Black scalebars are 100 nm long.} \label{figACFCompare}
\end{center}
\vspace{-0.9cm}
\end{figure*}

We can see the results of both methods in Fig.\,\ref{figACFCompare}. We take a simple vortex lattice image taken in Bi$_2$Pd from Ref.\,\cite{PhysRevB.94.014517}. There is experimental noise in the image and vortices appear as dots that are blurred. The statistical correlation is in essence the size of a vortex. We expect that $ACF(r)$ decreases with $r$, with a length scale that is the vortex size, and becomes negative at the intervortex distance.

We produce three columns where we have arbitrarily moved the colorscale zero vale. At one side of the histogram containing the values of the pixels (Fig.\,\ref{figACFCompare}(a,d,g)), at the center of the pixel values (Fig.\,\ref{figACFCompare}(b,e,h)) and at the average pixel of the histogram (Fig.\,\ref{figACFCompare}(c,f,i)).

In that way, we see that the result of the calculation is independent of the choice (Fig.\,\ref{figACFCompare}(d,e,f)) only when we use the statistical $ACF(r)$. As expected, we can determine the vortex core size from the decay at small r, and find a value which coincides with those provided in literature by analysing the shape of isolated vortices \cite{PhysRevB.94.014517}. The $ACF(r)$ becomes negative at the intervortex distance and oscillates, with a spatially decaying amplitude. The latter is due to the ratio of regions with zero DOS with respect to those with a finite DOS. The former is particularly large in this system\cite{PhysRevB.94.014517}.

By contrast, the computationally less demanding method of multiplying matrices only leads to a similar result when the zero of the histogram of the values in the image is centered at the average of the histogram (Fig.\,\ref{figACFCompare}(i)). Furthermore, the obtained length scale is strongly distorted.

Of course, a multiplication of 2D images also leads to a 2D image Fig.\,\ref{figACFCompare}(g,h,i). When using a statistical $ACF$, the result just depends on the polar coordinates $(r,\phi)$. In images that show no in-plane symmetry, as those discussed in the text, there is no angular dependence either and everything is in the radial dependence. When treating a vortex image with a six-fold symmetry, the radial dependence is as shown in the insets of Fig.\,\ref{figACFCompare}(d,e,f). The angular dependence is given however as a function of a single coordinate, $\phi$. For the purpose of comparison, we have streched this $\phi$ dependence into a 2D matrix in Figs.\,\ref{figACFCompare}(d,e,f). The relevant parameter is, though, the radial dependence of the correlations (insets of Figs.\,\ref{figACFCompare}(d-i)).

\section{Note D: Aspects of STM in high resistance films}

Let us remind a few basic aspects of tunneling STM experiments. In a usual STM experiment, there is often a small resistance in series with the tunnel junction. This helps controlling the noise level by producing low-pass filters together with capacitors. The tunneling current is given by

\begin{equation}
I (V)= V_0 R + V_1 \frac{1}{\sigma_{Junction}(V_1)}
\label{QeT1}
\end{equation}

where $R$ is the resistance in series, $V_0$ the voltage drop in that resistance, $V_1$ the voltage drop at the tip-sample junction and $\sigma_{Junction}$ the actual conductance of the junction. Usually, $\frac{1}{\sigma_{Junction}}\gg R$, so that $V_0\approx 0$ and $V\approx V_1$ and we can write the widely used relation between the tunneling current in a junction and the DOS:

\begin{equation}
I_t(V) \propto\int_0^{eV}dE N(E-eV)
\label{QeT2}
\end{equation}

Its derivative is $\sigma_{Junction}(V)=\frac{dI}{dV}$ is zero for $V<\Delta$, with $\Delta$ being the superconducting gap, finite $V>\Delta$ and eventually diverges exactly at the quasiparticle peaks for a conventional s-wave BCS superconductor. In our case, $R\approx$ 20k$\Omega$, and $\frac{1}{\sigma_{Junction}}$ is well above a M$\Omega$, so that the condition $\frac{1}{\sigma_{Junction}}\gg R$ is maintained for every purpose during the voltage ramp.

However, the current flows through the sample before reaching the tip-sample junction. Thus, there is an additional resistance $R_S$ which adds to $R$ due to the sample. It is important to see that the value of $R_S$ is the one found in macroscopic transport experiments and does not considerably modify $R$. For example, a sample resistance of 10 k$\Omega$ with a current flow of 1 nA leads to a voltage drop of 10 $\mu$V. Thus, the effect of $R+R_S$ is of at most a voltage shift of a few tens of $\mu V$. If we assume a non-ohmic $R_S$, the voltage drop in the sample might become larger. However, by varying the tip-sample distance we can modify $\sigma_{Junction}$, and thus the relative role of $R_S$ in the tunneling experiment. With a voltage drop in the sample, we expect to see shifts of features in the DOS of the sample. We do not observe such shifts, showing that $R_S$ is ohmic.

On the other hand, a DOS strongly varying with energy has further consequences that we need to discuss. To see this, we remind that the tip is positioned through a feedback mechanism that maintains a constant tunneling current at a bias of several mV, which is usually of order of a nA. The feedback mechanism thus imposes a relation between $I_t(V)$ which can be written as

\begin{equation}
I_t(V) \propto \frac{\int_0^{eV}dE N(E-eV)}{\int_0^{eV_B}dE N(E-eV)}
\label{QeT2}
\end{equation}

where $V_B$ is the bias voltage at which the control system of the STM is working. Thus, when modifying the bias voltage $V_{B1}>V_B$, $I_t(V)$ is modified by a factor that depends on the DOS integrated between the Fermi level and $V_B$. Its derivative is accordingly modified too. However, normalizing the current or its derivative at, say $V_0<V_B$, eliminates this factor.

\begin{equation}
\begin{aligned}
\frac{I_t(V)}{I_t(V_0)} &\propto \frac{\int_0^{eV}dE N(E-eV)}{\int_0^{eV_B}dE N(E-eV)} \frac{\int_0^{eV_B}dE N(E-eV)}{\int_0^{eV_0}dE N(E-eV)} \\
 &\propto  \frac{\int_0^{eV}dE N(E-eV)}{\int_0^{eV_0}dE N(E-eV)}
\label{QeT3}
\end{aligned}
\end{equation}

and provides curves that are comparable to each other with different bias voltages.

In the case we consider here, $I_t(V_1)$ is linked to the DOS, but we measure $I_t(V)$. The feedback mechanism acts in the same way, with modified normalization constants, which are eliminated by normalizing the current to $I$ at another bias voltage, $I_t(V_0)$. The same applies for the tunneling conductance $G(V)$. Although the influence of the feedback loop becomes more important in more resistive samples (as the heated D15), normalization at a fixed voltage can be used to obtain results that can be compared among samples and measurement conditions.

\bibliographystyle{apsrev4-1-titles}

\begin{thebibliography}{38}%
\makeatletter
\providecommand \@ifxundefined [1]{%
 \@ifx{#1\undefined}
}%
\providecommand \@ifnum [1]{%
 \ifnum #1\expandafter \@firstoftwo
 \else \expandafter \@secondoftwo
 \fi
}%
\providecommand \@ifx [1]{%
 \ifx #1\expandafter \@firstoftwo
 \else \expandafter \@secondoftwo
 \fi
}%
\providecommand \natexlab [1]{#1}%
\providecommand \enquote  [1]{``#1''}%
\providecommand \bibnamefont  [1]{#1}%
\providecommand \bibfnamefont [1]{#1}%
\providecommand \citenamefont [1]{#1}%
\providecommand \href@noop [0]{\@secondoftwo}%
\providecommand \href [0]{\begingroup \@sanitize@url \@href}%
\providecommand \@href[1]{\@@startlink{#1}\@@href}%
\providecommand \@@href[1]{\endgroup#1\@@endlink}%
\providecommand \@sanitize@url [0]{\catcode `\\12\catcode `\$12\catcode
  `\&12\catcode `\#12\catcode `\^12\catcode `\_12\catcode `\%12\relax}%
\providecommand \@@startlink[1]{}%
\providecommand \@@endlink[0]{}%
\providecommand \url  [0]{\begingroup\@sanitize@url \@url }%
\providecommand \@url [1]{\endgroup\@href {#1}{\urlprefix }}%
\providecommand \urlprefix  [0]{URL }%
\providecommand \Eprint [0]{\href }%
\providecommand \doibase [0]{http://doi.org/}%
\providecommand \selectlanguage [0]{\@gobble}%
\providecommand \bibinfo  [0]{\@secondoftwo}%
\providecommand \bibfield  [0]{\@secondoftwo}%
\providecommand \translation [1]{[#1]}%
\providecommand \BibitemOpen [0]{}%
\providecommand \bibitemStop [0]{}%
\providecommand \bibitemNoStop [0]{.\EOS\space}%
\providecommand \EOS [0]{\spacefactor3000\relax}%
\providecommand \BibitemShut  [1]{\csname bibitem#1\endcsname}%
\let\auto@bib@innerbib\@empty
\bibitem [{\citenamefont {Goldman}\ and\ \citenamefont
  {Markovic}(1998)}]{GoldmanReview}%
  \BibitemOpen
  \bibfield  {author} {\bibinfo {author} {\bibfnamefont {A.}~\bibnamefont
  {Goldman}}\ and\ \bibinfo {author} {\bibfnamefont {N.}~\bibnamefont
  {Markovic}},\ }\bibfield  {title} {\emph {\bibinfo {title} {Superconductor
  insulator transitions in the two dimensional limit}},\ }\href {\doibase
  10.1063/1.882069} {\bibfield  {journal} {\bibinfo  {journal} {Physics Today}\
  }\textbf {\bibinfo {volume} {51}},\ \bibinfo {pages} {39} (\bibinfo {year}
  {1998})}\BibitemShut {NoStop}%
\bibitem [{\citenamefont {Lin}\ \emph {et~al.}(2015)\citenamefont {Lin},
  \citenamefont {Nelson},\ and\ \citenamefont {Goldman}}]{LIN2015130}%
  \BibitemOpen
  \bibfield  {author} {\bibinfo {author} {\bibfnamefont {Y.-H.}\ \bibnamefont
  {Lin}}, \bibinfo {author} {\bibfnamefont {J.}~\bibnamefont {Nelson}}, \ and\
  \bibinfo {author} {\bibfnamefont {A.}~\bibnamefont {Goldman}},\ }\bibfield
  {title} {\emph {\bibinfo {title} {Superconductivity of very thin films: The
  superconductor insulator transition}},\ }\href {\doibase
  https://doi.org/10.1016/j.physc.2015.01.005} {\bibfield  {journal} {\bibinfo
  {journal} {Physica C: Superconductivity and its Applications}\ }\textbf
  {\bibinfo {volume} {514}},\ \bibinfo {pages} {130 } (\bibinfo {year}
  {2015})},\ \bibinfo {note} {superconducting Materials: Conventional,
  Unconventional and Undetermined}\BibitemShut {NoStop}%
\bibitem [{\citenamefont {Vinokur}\ and\ \citenamefont
  {et~al.}(2008)}]{Vinokur2008}%
  \BibitemOpen
  \bibfield  {author} {\bibinfo {author} {\bibfnamefont {V.}~\bibnamefont
  {Vinokur}}\ and\ \bibinfo {author} {\bibnamefont {et~al.}},\ }\bibfield
  {title} {\emph {\bibinfo {title} {Superinsulator and quantum
  synchronization}},\ }\href@noop {} {\bibfield  {journal} {\bibinfo  {journal}
  {Nature}\ }\textbf {\bibinfo {volume} {452}},\ \bibinfo {pages} {613}
  (\bibinfo {year} {2008})}\BibitemShut {NoStop}%
\bibitem [{\citenamefont {Kowal}\ and\ \citenamefont {Ovadyahu}(1994)}]{Zvi}%
  \BibitemOpen
  \bibfield  {author} {\bibinfo {author} {\bibfnamefont {D.}~\bibnamefont
  {Kowal}}\ and\ \bibinfo {author} {\bibfnamefont {Z.}~\bibnamefont
  {Ovadyahu}},\ }\bibfield  {title} {\emph {\bibinfo {title} {Disorder induced
  granularity in an amorphous superconductor}},\ }\href@noop {} {\bibfield
  {journal} {\bibinfo  {journal} {Solid St. Comm.}\ }\textbf {\bibinfo {volume}
  {90}},\ \bibinfo {pages} {783} (\bibinfo {year} {1994})}\BibitemShut
  {NoStop}%
\bibitem [{\citenamefont {Sac\'ep\'e}\ \emph {et~al.}(2008)\citenamefont
  {Sac\'ep\'e}, \citenamefont {Chapelier}, \citenamefont {Baturina},
  \citenamefont {Vinokur}, \citenamefont {Baklanov},\ and\ \citenamefont
  {Sanquer}}]{TiNInhom}%
  \BibitemOpen
  \bibfield  {author} {\bibinfo {author} {\bibfnamefont {B.}~\bibnamefont
  {Sac\'ep\'e}}, \bibinfo {author} {\bibfnamefont {C.}~\bibnamefont
  {Chapelier}}, \bibinfo {author} {\bibfnamefont {T.~I.}\ \bibnamefont
  {Baturina}}, \bibinfo {author} {\bibfnamefont {V.~M.}\ \bibnamefont
  {Vinokur}}, \bibinfo {author} {\bibfnamefont {M.~R.}\ \bibnamefont
  {Baklanov}}, \ and\ \bibinfo {author} {\bibfnamefont {M.}~\bibnamefont
  {Sanquer}},\ }\bibfield  {title} {\emph {\bibinfo {title} {Disorder-induced
  inhomogeneities of the superconducting state close to the
  superconductor-insulator transition}},\ }\href {\doibase
  10.1103/PhysRevLett.101.157006} {\bibfield  {journal} {\bibinfo  {journal}
  {Phys. Rev. Lett.}\ }\textbf {\bibinfo {volume} {101}},\ \bibinfo {pages}
  {157006} (\bibinfo {year} {2008})}\BibitemShut {NoStop}%
\bibitem [{\citenamefont {Sherman}\ \emph {et~al.}(2012)\citenamefont
  {Sherman}, \citenamefont {Kopnov}, \citenamefont {Shahar},\ and\
  \citenamefont {Frydman}}]{Shahar}%
  \BibitemOpen
  \bibfield  {author} {\bibinfo {author} {\bibfnamefont {D.}~\bibnamefont
  {Sherman}}, \bibinfo {author} {\bibfnamefont {G.}~\bibnamefont {Kopnov}},
  \bibinfo {author} {\bibfnamefont {D.}~\bibnamefont {Shahar}}, \ and\ \bibinfo
  {author} {\bibfnamefont {A.}~\bibnamefont {Frydman}},\ }\bibfield  {title}
  {\emph {\bibinfo {title} {Measurement of a superconducting energy gap in a
  homogeneously amorphous insulator}},\ }\href@noop {} {\bibfield  {journal}
  {\bibinfo  {journal} {Phys Rev Lett}\ }\textbf {\bibinfo {volume} {108}},\
  \bibinfo {pages} {177006} (\bibinfo {year} {2012})}\BibitemShut {NoStop}%
\bibitem [{\citenamefont {B.Sacepe}\ \emph {et~al.}(2011)\citenamefont
  {B.Sacepe}, \citenamefont {T.Dubouchet}, \citenamefont {Chapelier},
  \citenamefont {M.Sanquer}, \citenamefont {Ovadia}, \citenamefont {Shahar},
  \citenamefont {Feigelman},\ and\ \citenamefont {Ioffe}}]{NoPeak}%
  \BibitemOpen
  \bibfield  {author} {\bibinfo {author} {\bibnamefont {B.Sacepe}}, \bibinfo
  {author} {\bibnamefont {T.Dubouchet}}, \bibinfo {author} {\bibfnamefont
  {C.}~\bibnamefont {Chapelier}}, \bibinfo {author} {\bibnamefont {M.Sanquer}},
  \bibinfo {author} {\bibfnamefont {M.}~\bibnamefont {Ovadia}}, \bibinfo
  {author} {\bibfnamefont {D.}~\bibnamefont {Shahar}}, \bibinfo {author}
  {\bibfnamefont {M.}~\bibnamefont {Feigelman}}, \ and\ \bibinfo {author}
  {\bibfnamefont {L.}~\bibnamefont {Ioffe}},\ }\bibfield  {title} {\emph
  {\bibinfo {title} {Localization of preformed cooper pairs in disordered
  superconductors}},\ }\href@noop {} {\bibfield  {journal} {\bibinfo  {journal}
  {Nature Phys.}\ }\textbf {\bibinfo {volume} {7}},\ \bibinfo {pages} {239}
  (\bibinfo {year} {2011})}\BibitemShut {NoStop}%
\bibitem [{\citenamefont {Sacepe}\ \emph {et~al.}(2010)\citenamefont {Sacepe},
  \citenamefont {Chapelier}, \citenamefont {Baturina}, \citenamefont {Vinokur},
  \citenamefont {Baklanov},\ and\ \citenamefont {M.Sanquer}}]{TiNPG}%
  \BibitemOpen
  \bibfield  {author} {\bibinfo {author} {\bibfnamefont {B.}~\bibnamefont
  {Sacepe}}, \bibinfo {author} {\bibfnamefont {C.}~\bibnamefont {Chapelier}},
  \bibinfo {author} {\bibfnamefont {T.}~\bibnamefont {Baturina}}, \bibinfo
  {author} {\bibfnamefont {V.}~\bibnamefont {Vinokur}}, \bibinfo {author}
  {\bibfnamefont {M.}~\bibnamefont {Baklanov}}, \ and\ \bibinfo {author}
  {\bibnamefont {M.Sanquer}},\ }\bibfield  {title} {\emph {\bibinfo {title}
  {Pseudogap in a thin film of a conventional superconductor}},\ }\href@noop {}
  {\bibfield  {journal} {\bibinfo  {journal} {Nature Comms.}\ }\textbf
  {\bibinfo {volume} {1}},\ \bibinfo {pages} {140} (\bibinfo {year}
  {2010})}\BibitemShut {NoStop}%
\bibitem [{\citenamefont {Mondal}\ \emph {et~al.}(2011)\citenamefont {Mondal},
  \citenamefont {Kamlapure}, \citenamefont {Chand}, \citenamefont {Saraswat},
  \citenamefont {Kumar}, \citenamefont {Jesudasan}, \citenamefont {Benfatto},
  \citenamefont {Tripathi},\ and\ \citenamefont {Raychaudhuri}}]{PratapPRL11}%
  \BibitemOpen
  \bibfield  {author} {\bibinfo {author} {\bibfnamefont {M.}~\bibnamefont
  {Mondal}}, \bibinfo {author} {\bibfnamefont {A.}~\bibnamefont {Kamlapure}},
  \bibinfo {author} {\bibfnamefont {M.}~\bibnamefont {Chand}}, \bibinfo
  {author} {\bibfnamefont {G.}~\bibnamefont {Saraswat}}, \bibinfo {author}
  {\bibfnamefont {S.}~\bibnamefont {Kumar}}, \bibinfo {author} {\bibfnamefont
  {J.}~\bibnamefont {Jesudasan}}, \bibinfo {author} {\bibfnamefont
  {L.}~\bibnamefont {Benfatto}}, \bibinfo {author} {\bibfnamefont
  {V.}~\bibnamefont {Tripathi}}, \ and\ \bibinfo {author} {\bibfnamefont
  {P.}~\bibnamefont {Raychaudhuri}},\ }\bibfield  {title} {\emph {\bibinfo
  {title} {Phase fluctuations in a stongly disordered s-wave {NbN}
  superconductor close to the metal-insulator transition}},\ }\href@noop {}
  {\bibfield  {journal} {\bibinfo  {journal} {Phys Rev Lett}\ }\textbf
  {\bibinfo {volume} {106}},\ \bibinfo {pages} {047001} (\bibinfo {year}
  {2011})}\BibitemShut {NoStop}%
\bibitem [{\citenamefont {Kalampure}\ \emph {et~al.}(2013)\citenamefont
  {Kalampure}, \citenamefont {Das}, \citenamefont {Ganguli}, \citenamefont
  {Parmar}, \citenamefont {Bhattacharyya},\ and\ \citenamefont
  {Raydchahuri}}]{PratapScirep13}%
  \BibitemOpen
  \bibfield  {author} {\bibinfo {author} {\bibfnamefont {A.}~\bibnamefont
  {Kalampure}}, \bibinfo {author} {\bibfnamefont {T.}~\bibnamefont {Das}},
  \bibinfo {author} {\bibfnamefont {C.}~\bibnamefont {Ganguli}}, \bibinfo
  {author} {\bibfnamefont {J.}~\bibnamefont {Parmar}}, \bibinfo {author}
  {\bibfnamefont {S.}~\bibnamefont {Bhattacharyya}}, \ and\ \bibinfo {author}
  {\bibfnamefont {P.}~\bibnamefont {Raydchahuri}},\ }\bibfield  {title} {\emph
  {\bibinfo {title} {Emergence of nanoscale inhomogeneity in the
  superconducting state of a homogeneously disordered conventional
  superconductor}},\ }\href@noop {} {\bibfield  {journal} {\bibinfo  {journal}
  {Sci. Rep.}\ }\textbf {\bibinfo {volume} {3}},\ \bibinfo {pages} {2979}
  (\bibinfo {year} {2013})}\BibitemShut {NoStop}%
\bibitem [{\citenamefont {Altshuler}\ and\ \citenamefont {Aronov}(1985)}]{AA}%
  \BibitemOpen
  \bibfield  {author} {\bibinfo {author} {\bibfnamefont {B.}~\bibnamefont
  {Altshuler}}\ and\ \bibinfo {author} {\bibfnamefont {A.}~\bibnamefont
  {Aronov}},\ }\bibfield  {title} {\emph {\bibinfo {title} {Electron electron
  interaction in disordered conductors}},\ }\href@noop {} {\bibfield  {journal}
  {\bibinfo  {journal} {Modern Problems in Condensed Matter Sciences}\ }\textbf
  {\bibinfo {volume} {10}},\ \bibinfo {pages} {1} (\bibinfo {year}
  {1985})}\BibitemShut {NoStop}%
\bibitem [{\citenamefont {Valles}\ \emph {et~al.}(1989)\citenamefont {Valles},
  \citenamefont {Dynes},\ and\ \citenamefont {Garno}}]{Valles89}%
  \BibitemOpen
  \bibfield  {author} {\bibinfo {author} {\bibfnamefont {J.~M.}\ \bibnamefont
  {Valles}}, \bibinfo {author} {\bibfnamefont {R.~C.}\ \bibnamefont {Dynes}}, \
  and\ \bibinfo {author} {\bibfnamefont {J.~P.}\ \bibnamefont {Garno}},\
  }\bibfield  {title} {\emph {\bibinfo {title} {Temperature dependence of the
  two-dimensional electronic density of states in disordered metal films}},\
  }\href {\doibase 10.1103/PhysRevB.40.7590} {\bibfield  {journal} {\bibinfo
  {journal} {Phys. Rev. B}\ }\textbf {\bibinfo {volume} {40}},\ \bibinfo
  {pages} {7590} (\bibinfo {year} {1989})}\BibitemShut {NoStop}%
\bibitem [{\citenamefont {Butko}\ \emph {et~al.}(2000)\citenamefont {Butko},
  \citenamefont {DiTusa},\ and\ \citenamefont {Adams}}]{Butko2000}%
  \BibitemOpen
  \bibfield  {author} {\bibinfo {author} {\bibfnamefont {V.~Y.}\ \bibnamefont
  {Butko}}, \bibinfo {author} {\bibfnamefont {J.~F.}\ \bibnamefont {DiTusa}}, \
  and\ \bibinfo {author} {\bibfnamefont {P.~W.}\ \bibnamefont {Adams}},\
  }\bibfield  {title} {\emph {\bibinfo {title} {Coulomb gap: How a metal film
  becomes an insulator}},\ }\href {\doibase 10.1103/PhysRevLett.84.1543}
  {\bibfield  {journal} {\bibinfo  {journal} {Phys. Rev. Lett.}\ }\textbf
  {\bibinfo {volume} {84}},\ \bibinfo {pages} {1543} (\bibinfo {year}
  {2000})}\BibitemShut {NoStop}%
\bibitem [{\citenamefont {Bielejec}\ \emph {et~al.}(2001)\citenamefont
  {Bielejec}, \citenamefont {Ruan},\ and\ \citenamefont {Wu}}]{BeDOS}%
  \BibitemOpen
  \bibfield  {author} {\bibinfo {author} {\bibfnamefont {E.}~\bibnamefont
  {Bielejec}}, \bibinfo {author} {\bibfnamefont {J.}~\bibnamefont {Ruan}}, \
  and\ \bibinfo {author} {\bibfnamefont {W.}~\bibnamefont {Wu}},\ }\bibfield
  {title} {\emph {\bibinfo {title} {Hard correlation gap observed in
  quench-condensed ultrathin beryllium}},\ }\href@noop {} {\bibfield  {journal}
  {\bibinfo  {journal} {Phys. Rev. Lett.}\ }\textbf {\bibinfo {volume} {87}},\
  \bibinfo {pages} {036801} (\bibinfo {year} {2001})}\BibitemShut {NoStop}%
\bibitem [{\citenamefont {Kopietz}(1998)}]{BK0}%
  \BibitemOpen
  \bibfield  {author} {\bibinfo {author} {\bibfnamefont {P.}~\bibnamefont
  {Kopietz}},\ }\bibfield  {title} {\emph {\bibinfo {title} {Coulomb gap in the
  density of states of disordered metals in two dimensions}},\ }\href {\doibase
  10.1103/PhysRevLett.81.2120} {\bibfield  {journal} {\bibinfo  {journal}
  {Phys. Rev. Lett.}\ }\textbf {\bibinfo {volume} {81}},\ \bibinfo {pages}
  {2120} (\bibinfo {year} {1998})}\BibitemShut {NoStop}%
\bibitem [{\citenamefont {Bartosch}\ and\ \citenamefont {Kopietz}(2002)}]{BK}%
  \BibitemOpen
  \bibfield  {author} {\bibinfo {author} {\bibfnamefont {L.}~\bibnamefont
  {Bartosch}}\ and\ \bibinfo {author} {\bibfnamefont {P.}~\bibnamefont
  {Kopietz}},\ }\bibfield  {title} {\emph {\bibinfo {title} {Zero bias anomaly
  in the density of states of low-dimensional metals}},\ }\href {\doibase
  10.1140/epjb/e2002-00210-2} {\bibfield  {journal} {\bibinfo  {journal} {The
  European Physical Journal B - Condensed Matter and Complex Systems}\ }\textbf
  {\bibinfo {volume} {28}},\ \bibinfo {pages} {29} (\bibinfo {year}
  {2002})}\BibitemShut {NoStop}%
\bibitem [{\citenamefont {Rodrigo}\ \emph {et~al.}(2004)\citenamefont
  {Rodrigo}, \citenamefont {Suderow},\ and\ \citenamefont
  {Vieira}}]{Rodrigo04}%
  \BibitemOpen
  \bibfield  {author} {\bibinfo {author} {\bibfnamefont {J.~G.}\ \bibnamefont
  {Rodrigo}}, \bibinfo {author} {\bibfnamefont {H.}~\bibnamefont {Suderow}}, \
  and\ \bibinfo {author} {\bibfnamefont {S.}~\bibnamefont {Vieira}},\
  }\bibfield  {title} {\emph {\bibinfo {title} {On the use of {STM}
  superconducting tips at very low temperatures}},\ }\href@noop {} {\bibfield
  {journal} {\bibinfo  {journal} {European Phys. Journal B}\ }\textbf {\bibinfo
  {volume} {40}},\ \bibinfo {pages} {483} (\bibinfo {year} {2004})}\BibitemShut
  {NoStop}%
\bibitem [{\citenamefont {Guillam\'on}\ \emph {et~al.}(2008)\citenamefont
  {Guillam\'on}, \citenamefont {Suderow}, \citenamefont {Vieira},\ and\
  \citenamefont {Rodiere}}]{Guillamon07}%
  \BibitemOpen
  \bibfield  {author} {\bibinfo {author} {\bibfnamefont {I.}~\bibnamefont
  {Guillam\'on}}, \bibinfo {author} {\bibfnamefont {H.}~\bibnamefont
  {Suderow}}, \bibinfo {author} {\bibfnamefont {S.}~\bibnamefont {Vieira}}, \
  and\ \bibinfo {author} {\bibfnamefont {P.}~\bibnamefont {Rodiere}},\
  }\bibfield  {title} {\emph {\bibinfo {title} {Scanning tunneling spectroscopy
  with superconducting tips of al}},\ }\href@noop {} {\bibfield  {journal}
  {\bibinfo  {journal} {Physica C}\ }\textbf {\bibinfo {volume} {468}},\
  \bibinfo {pages} {537} (\bibinfo {year} {2008})}\BibitemShut {NoStop}%
\bibitem [{\citenamefont {Suderow}\ \emph {et~al.}(2011)\citenamefont
  {Suderow}, \citenamefont {Guillamon},\ and\ \citenamefont
  {Vieira}}]{Suderow11}%
  \BibitemOpen
  \bibfield  {author} {\bibinfo {author} {\bibfnamefont {H.}~\bibnamefont
  {Suderow}}, \bibinfo {author} {\bibfnamefont {I.}~\bibnamefont {Guillamon}},
  \ and\ \bibinfo {author} {\bibfnamefont {S.}~\bibnamefont {Vieira}},\
  }\bibfield  {title} {\emph {\bibinfo {title} {Compact very low temperature
  scanning tunneling microscope with mechanically driven horizontal linear
  positioning stage.}},\ }\href@noop {} {\bibfield  {journal} {\bibinfo
  {journal} {Rev. Sci. Inst.}\ }\textbf {\bibinfo {volume} {82}},\ \bibinfo
  {pages} {033711} (\bibinfo {year} {2011})}\BibitemShut {NoStop}%
\bibitem [{\citenamefont {Baturina}\ \emph {et~al.}(2012)\citenamefont
  {Baturina}, \citenamefont {Postolova}, \citenamefont {Mironov}, \citenamefont
  {A.~Glatz},\ and\ \citenamefont {Vinokur}}]{Baturina12EPL}%
  \BibitemOpen
  \bibfield  {author} {\bibinfo {author} {\bibfnamefont {T.~I.}\ \bibnamefont
  {Baturina}}, \bibinfo {author} {\bibfnamefont {S.}~\bibnamefont {Postolova}},
  \bibinfo {author} {\bibfnamefont {A.}~\bibnamefont {Mironov}}, \bibinfo
  {author} {\bibfnamefont {M.~B.}\ \bibnamefont {A.~Glatz}}, \ and\ \bibinfo
  {author} {\bibfnamefont {V.}~\bibnamefont {Vinokur}},\ }\bibfield  {title}
  {\emph {\bibinfo {title} {Superconducting phase transitions in ultrathin
  {TiN} films}},\ }\href@noop {} {\bibfield  {journal} {\bibinfo  {journal}
  {Europhys. Lett.}\ }\textbf {\bibinfo {volume} {97}},\ \bibinfo {pages}
  {17012} (\bibinfo {year} {2012})}\BibitemShut {NoStop}%
\bibitem [{\citenamefont {Hager}\ \emph {et~al.}(2005)\citenamefont {Hager},
  \citenamefont {Matzdorf}, \citenamefont {He}, \citenamefont {Jin},
  \citenamefont {Mandrus}, \citenamefont {Cazalilla},\ and\ \citenamefont
  {Plummer}}]{1dBK1}%
  \BibitemOpen
  \bibfield  {author} {\bibinfo {author} {\bibfnamefont {J.}~\bibnamefont
  {Hager}}, \bibinfo {author} {\bibfnamefont {R.}~\bibnamefont {Matzdorf}},
  \bibinfo {author} {\bibfnamefont {J.}~\bibnamefont {He}}, \bibinfo {author}
  {\bibfnamefont {R.}~\bibnamefont {Jin}}, \bibinfo {author} {\bibfnamefont
  {D.}~\bibnamefont {Mandrus}}, \bibinfo {author} {\bibfnamefont {M.~A.}\
  \bibnamefont {Cazalilla}}, \ and\ \bibinfo {author} {\bibfnamefont {E.~W.}\
  \bibnamefont {Plummer}},\ }\bibfield  {title} {\emph {\bibinfo {title}
  {Non-fermi-liquid behavior in quasi-one-dimensional
  ${\mathrm{li}}_{0.9}{\mathrm{mo}}_{6}{\mathrm{o}}_{17}$}},\ }\href {\doibase
  10.1103/PhysRevLett.95.186402} {\bibfield  {journal} {\bibinfo  {journal}
  {Phys. Rev. Lett.}\ }\textbf {\bibinfo {volume} {95}},\ \bibinfo {pages}
  {186402} (\bibinfo {year} {2005})}\BibitemShut {NoStop}%
\bibitem [{\citenamefont {Blumenstein}\ \emph {et~al.}(2011)\citenamefont
  {Blumenstein}, \citenamefont {Sch{\"a}fer}, \citenamefont {Mietke},
  \citenamefont {Meyer}, \citenamefont {Dollinger}, \citenamefont {Lochner},
  \citenamefont {Cui}, \citenamefont {Patthey}, \citenamefont {Matzdorf},\ and\
  \citenamefont {Claessen}}]{1dBK2}%
  \BibitemOpen
  \bibfield  {author} {\bibinfo {author} {\bibfnamefont {C.}~\bibnamefont
  {Blumenstein}}, \bibinfo {author} {\bibfnamefont {J.}~\bibnamefont
  {Sch{\"a}fer}}, \bibinfo {author} {\bibfnamefont {S.}~\bibnamefont {Mietke}},
  \bibinfo {author} {\bibfnamefont {S.}~\bibnamefont {Meyer}}, \bibinfo
  {author} {\bibfnamefont {A.}~\bibnamefont {Dollinger}}, \bibinfo {author}
  {\bibfnamefont {M.}~\bibnamefont {Lochner}}, \bibinfo {author} {\bibfnamefont
  {X.~Y.}\ \bibnamefont {Cui}}, \bibinfo {author} {\bibfnamefont
  {L.}~\bibnamefont {Patthey}}, \bibinfo {author} {\bibfnamefont
  {R.}~\bibnamefont {Matzdorf}}, \ and\ \bibinfo {author} {\bibfnamefont
  {R.}~\bibnamefont {Claessen}},\ }\bibfield  {title} {\emph {\bibinfo {title}
  {Atomically controlled quantum chains hosting a tomonaga-luttinger liquid}},\
  }\href {\doibase 10.1038/nphys2051} {\bibfield  {journal} {\bibinfo
  {journal} {Nature Physics}\ }\textbf {\bibinfo {volume} {7}},\ \bibinfo
  {pages} {776} (\bibinfo {year} {2011})}\BibitemShut {NoStop}%
\bibitem [{\citenamefont {Yan}\ \emph {et~al.}(2012)\citenamefont {Yan},
  \citenamefont {Xu}, \citenamefont {Hong}, \citenamefont {Sun}, \citenamefont
  {Feng}, \citenamefont {Nie},\ and\ \citenamefont {He}}]{1dBK3}%
  \BibitemOpen
  \bibfield  {author} {\bibinfo {author} {\bibfnamefont {H.}~\bibnamefont
  {Yan}}, \bibinfo {author} {\bibfnamefont {R.}~\bibnamefont {Xu}}, \bibinfo
  {author} {\bibfnamefont {X.}~\bibnamefont {Hong}}, \bibinfo {author}
  {\bibfnamefont {Y.}~\bibnamefont {Sun}}, \bibinfo {author} {\bibfnamefont
  {L.}~\bibnamefont {Feng}}, \bibinfo {author} {\bibfnamefont {J.-C.}\
  \bibnamefont {Nie}}, \ and\ \bibinfo {author} {\bibfnamefont
  {L.}~\bibnamefont {He}},\ }\bibfield  {title} {\emph {\bibinfo {title}
  {Zero-bias anomaly in one-dimensional ultrathin metallic nanowires}},\ }\href
  {\doibase 10.1063/1.4747799} {\bibfield  {journal} {\bibinfo  {journal} {AIP
  Advances}\ }\textbf {\bibinfo {volume} {2}},\ \bibinfo {pages} {032143}
  (\bibinfo {year} {2012})},\ \Eprint
  {http://arxiv.org/abs/https://doi.org/10.1063/1.4747799}
  {https://doi.org/10.1063/1.4747799} \BibitemShut {NoStop}%
\bibitem [{\citenamefont {Finkel'stein}(1994)}]{Fink}%
  \BibitemOpen
  \bibfield  {author} {\bibinfo {author} {\bibfnamefont {A.}~\bibnamefont
  {Finkel'stein}},\ }\bibfield  {title} {\emph {\bibinfo {title} {Suppression
  of superconductivity in homogeneously disordered systems}},\ }\href {\doibase
  https://doi.org/10.1016/0921-4526(94)90267-4} {\bibfield  {journal} {\bibinfo
   {journal} {Physica B: Condensed Matter}\ }\textbf {\bibinfo {volume}
  {197}},\ \bibinfo {pages} {636 } (\bibinfo {year} {1994})}\BibitemShut
  {NoStop}%
\bibitem [{\citenamefont {Postolova}\ \emph {et~al.}(2017)\citenamefont
  {Postolova}, \citenamefont {Mironov}, \citenamefont {Baklanov}, \citenamefont
  {Vinokur},\ and\ \citenamefont {Baturina}}]{Postolova2017}%
  \BibitemOpen
  \bibfield  {author} {\bibinfo {author} {\bibfnamefont {S.~V.}\ \bibnamefont
  {Postolova}}, \bibinfo {author} {\bibfnamefont {A.~Y.}\ \bibnamefont
  {Mironov}}, \bibinfo {author} {\bibfnamefont {M.~R.}\ \bibnamefont
  {Baklanov}}, \bibinfo {author} {\bibfnamefont {V.~M.}\ \bibnamefont
  {Vinokur}}, \ and\ \bibinfo {author} {\bibfnamefont {T.~I.}\ \bibnamefont
  {Baturina}},\ }\bibfield  {title} {\emph {\bibinfo {title} {Reentrant
  resistive behavior and dimensional crossover in disordered superconducting
  tin films}},\ }\href {\doibase 10.1038/s41598-017-01753-w} {\bibfield
  {journal} {\bibinfo  {journal} {Scientific Reports}\ }\textbf {\bibinfo
  {volume} {7}},\ \bibinfo {pages} {1718} (\bibinfo {year} {2017})}\BibitemShut
  {NoStop}%
\bibitem [{\citenamefont {Khomskii}\ and\ \citenamefont
  {Kustmartsev}(1992)}]{dRo}%
  \BibitemOpen
  \bibfield  {author} {\bibinfo {author} {\bibfnamefont {D.}~\bibnamefont
  {Khomskii}}\ and\ \bibinfo {author} {\bibfnamefont {F.}~\bibnamefont
  {Kustmartsev}},\ }\bibfield  {title} {\emph {\bibinfo {title} {Charge
  redistribution and properties of high-temperature superconductors}},\
  }\href@noop {} {\bibfield  {journal} {\bibinfo  {journal} {Phys Rev B}\
  }\textbf {\bibinfo {volume} {46}},\ \bibinfo {pages} {14245} (\bibinfo {year}
  {1992})}\BibitemShut {NoStop}%
\bibitem [{\citenamefont {Fratini}\ \emph {et~al.}(2010)\citenamefont
  {Fratini}, \citenamefont {Poccia}, \citenamefont {Ricci}, \citenamefont
  {Campi}, \citenamefont {Burghammer}, \citenamefont {Aeppli},\ and\
  \citenamefont {Bianconi}}]{Fratini2010}%
  \BibitemOpen
  \bibfield  {author} {\bibinfo {author} {\bibfnamefont {M.}~\bibnamefont
  {Fratini}}, \bibinfo {author} {\bibfnamefont {N.}~\bibnamefont {Poccia}},
  \bibinfo {author} {\bibfnamefont {A.}~\bibnamefont {Ricci}}, \bibinfo
  {author} {\bibfnamefont {G.}~\bibnamefont {Campi}}, \bibinfo {author}
  {\bibfnamefont {M.}~\bibnamefont {Burghammer}}, \bibinfo {author}
  {\bibfnamefont {G.}~\bibnamefont {Aeppli}}, \ and\ \bibinfo {author}
  {\bibfnamefont {A.}~\bibnamefont {Bianconi}},\ }\bibfield  {title} {\emph
  {\bibinfo {title} {Scale-free structural organization of oxygen interstitials
  in {La$_2$CuO$_{4+y}$}}},\ }\href {\doibase 10.1038/nature09260} {\bibfield
  {journal} {\bibinfo  {journal} {Nature}\ }\textbf {\bibinfo {volume} {466}},\
  \bibinfo {pages} {841} (\bibinfo {year} {2010})}\BibitemShut {NoStop}%
\bibitem [{\citenamefont {Giraldo-Gallo}\ \emph {et~al.}(2015)\citenamefont
  {Giraldo-Gallo}, \citenamefont {Zhang}, \citenamefont {Parra}, \citenamefont
  {Manoharan}, \citenamefont {Beasley}, \citenamefont {Geballe}, \citenamefont
  {Kramer},\ and\ \citenamefont {Fisher}}]{Giraldo-Gallo2015}%
  \BibitemOpen
  \bibfield  {author} {\bibinfo {author} {\bibfnamefont {P.}~\bibnamefont
  {Giraldo-Gallo}}, \bibinfo {author} {\bibfnamefont {Y.}~\bibnamefont
  {Zhang}}, \bibinfo {author} {\bibfnamefont {C.}~\bibnamefont {Parra}},
  \bibinfo {author} {\bibfnamefont {H.~C.}\ \bibnamefont {Manoharan}}, \bibinfo
  {author} {\bibfnamefont {M.~R.}\ \bibnamefont {Beasley}}, \bibinfo {author}
  {\bibfnamefont {T.~H.}\ \bibnamefont {Geballe}}, \bibinfo {author}
  {\bibfnamefont {M.~J.}\ \bibnamefont {Kramer}}, \ and\ \bibinfo {author}
  {\bibfnamefont {I.~R.}\ \bibnamefont {Fisher}},\ }\bibfield  {title} {\emph
  {\bibinfo {title} {Stripe-like nanoscale structural phase separation in
  superconducting {BaPb1-xBixO3}}},\ }\href {\doibase 10.1038/ncomms9231}
  {\bibfield  {journal} {\bibinfo  {journal} {Nature Communications}\ }\textbf
  {\bibinfo {volume} {6}},\ \bibinfo {pages} {8231} (\bibinfo {year}
  {2015})}\BibitemShut {NoStop}%
\bibitem [{\citenamefont {Neckel}\ \emph {et~al.}(1975)\citenamefont {Neckel},
  \citenamefont {Rastl}, \citenamefont {Eibler}, \citenamefont {Weinberger},\
  and\ \citenamefont {Schwarz}}]{Neckel_1975}%
  \BibitemOpen
  \bibfield  {author} {\bibinfo {author} {\bibfnamefont {A.}~\bibnamefont
  {Neckel}}, \bibinfo {author} {\bibfnamefont {P.}~\bibnamefont {Rastl}},
  \bibinfo {author} {\bibfnamefont {R.}~\bibnamefont {Eibler}}, \bibinfo
  {author} {\bibfnamefont {P.}~\bibnamefont {Weinberger}}, \ and\ \bibinfo
  {author} {\bibfnamefont {K.}~\bibnamefont {Schwarz}},\ }\bibfield  {title}
  {\emph {\bibinfo {title} {Results of self-consistent band-structure
  calculations for {ScN}, {ScO}, {TiC}, {TiN}, {TiO}, {VC}, {VN} and {VO}}},\
  }\href {\doibase 10.1088/0022-3719/9/4/008} {\bibfield  {journal} {\bibinfo
  {journal} {Journal of Physics C: Solid State Physics}\ }\textbf {\bibinfo
  {volume} {9}},\ \bibinfo {pages} {579} (\bibinfo {year} {1975})}\BibitemShut
  {NoStop}%
\bibitem [{\citenamefont {Benito}\ and\ \citenamefont {et~al}()}]{Benito19}%
  \BibitemOpen
  \bibfield  {author} {\bibinfo {author} {\bibfnamefont {J.}~\bibnamefont
  {Benito}}\ and\ \bibinfo {author} {\bibnamefont {et~al}},\ }\href@noop {}
  {}\bibinfo {note} {Https://arxiv.org/abs/1904.10999}\BibitemShut {NoStop}%
\bibitem [{cod()}]{codeFractal}%
  \BibitemOpen
  \href@noop {} {}\bibinfo {note}
  {Https://github.com/LowTemperaturesUAM/MultiFractal-Analysis}\BibitemShut
  {NoStop}%
\bibitem [{\citenamefont {Feigel'man}\ \emph {et~al.}(2001)\citenamefont
  {Feigel'man}, \citenamefont {Ioffe}, \citenamefont {Kravtsov},\ and\
  \citenamefont {Yuzbashyan}}]{Feigelman07}%
  \BibitemOpen
  \bibfield  {author} {\bibinfo {author} {\bibfnamefont {M.}~\bibnamefont
  {Feigel'man}}, \bibinfo {author} {\bibfnamefont {M.}~\bibnamefont {Ioffe}},
  \bibinfo {author} {\bibfnamefont {L.}~\bibnamefont {Kravtsov}}, \ and\
  \bibinfo {author} {\bibfnamefont {V.}~\bibnamefont {Yuzbashyan}},\ }\bibfield
   {title} {\emph {\bibinfo {title} {Eigenfunction fractality and pseudogap
  state near the superconductor-insulator transition}},\ }\href@noop {}
  {\bibfield  {journal} {\bibinfo  {journal} {Phys. Rev. Lett.}\ }\textbf
  {\bibinfo {volume} {98}},\ \bibinfo {pages} {027001} (\bibinfo {year}
  {2001})}\BibitemShut {NoStop}%
\bibitem [{\citenamefont {Feigel'man}\ \emph {et~al.}(2010)\citenamefont
  {Feigel'man}, \citenamefont {Ioffe}, \citenamefont {Kravtsov},\ and\
  \citenamefont {Cuevas}}]{FEIGELMAN20101390}%
  \BibitemOpen
  \bibfield  {author} {\bibinfo {author} {\bibfnamefont {M.}~\bibnamefont
  {Feigel'man}}, \bibinfo {author} {\bibfnamefont {L.}~\bibnamefont {Ioffe}},
  \bibinfo {author} {\bibfnamefont {V.}~\bibnamefont {Kravtsov}}, \ and\
  \bibinfo {author} {\bibfnamefont {E.}~\bibnamefont {Cuevas}},\ }\bibfield
  {title} {\emph {\bibinfo {title} {Fractal superconductivity near localization
  threshold}},\ }\href {\doibase https://doi.org/10.1016/j.aop.2010.04.001}
  {\bibfield  {journal} {\bibinfo  {journal} {Annals of Physics}\ }\textbf
  {\bibinfo {volume} {325}},\ \bibinfo {pages} {1390 } (\bibinfo {year}
  {2010})},\ \bibinfo {note} {july 2010 Special Issue}\BibitemShut {NoStop}%
\bibitem [{\citenamefont {Carbillet}\ \emph {et~al.}(2019)\citenamefont
  {Carbillet}, \citenamefont {Cherkez}, \citenamefont {Skvortsov},
  \citenamefont {Feigel'man}, \citenamefont {Debontridder}, \citenamefont
  {Ioffe}, \citenamefont {Stolyarov}, \citenamefont {Ilin}, \citenamefont
  {Siegel}, \citenamefont {Roditchev}, \citenamefont {Cren},\ and\
  \citenamefont {Brun}}]{carbillet2019spectroscopic}%
  \BibitemOpen
  \bibfield  {author} {\bibinfo {author} {\bibfnamefont {C.}~\bibnamefont
  {Carbillet}}, \bibinfo {author} {\bibfnamefont {V.}~\bibnamefont {Cherkez}},
  \bibinfo {author} {\bibfnamefont {M.~A.}\ \bibnamefont {Skvortsov}}, \bibinfo
  {author} {\bibfnamefont {M.~V.}\ \bibnamefont {Feigel'man}}, \bibinfo
  {author} {\bibfnamefont {F.}~\bibnamefont {Debontridder}}, \bibinfo {author}
  {\bibfnamefont {L.~B.}\ \bibnamefont {Ioffe}}, \bibinfo {author}
  {\bibfnamefont {V.~S.}\ \bibnamefont {Stolyarov}}, \bibinfo {author}
  {\bibfnamefont {K.}~\bibnamefont {Ilin}}, \bibinfo {author} {\bibfnamefont
  {M.}~\bibnamefont {Siegel}}, \bibinfo {author} {\bibfnamefont
  {D.}~\bibnamefont {Roditchev}}, \bibinfo {author} {\bibfnamefont
  {T.}~\bibnamefont {Cren}}, \ and\ \bibinfo {author} {\bibfnamefont
  {C.}~\bibnamefont {Brun}},\ }\href@noop {} {\emph {\bibinfo {title}
  {Spectroscopic evidence for strong correlations between local resistance and
  superconducting gap in ultrathin {NbN} films}}} (\bibinfo {year} {2019}),\
  \Eprint {http://arxiv.org/abs/1903.01802} {arXiv:1903.01802
  [cond-mat.supr-con]} \BibitemShut {NoStop}%
\bibitem [{\citenamefont {Finkelstein}(1987)}]{Finkelstein}%
  \BibitemOpen
  \bibfield  {author} {\bibinfo {author} {\bibfnamefont {A.~M.}\ \bibnamefont
  {Finkelstein}},\ }\href@noop {} {\bibfield  {journal} {\bibinfo  {journal}
  {JETP Lett}\ }\textbf {\bibinfo {volume} {45}},\ \bibinfo {pages} {46}
  (\bibinfo {year} {1987})}\BibitemShut {NoStop}%
\bibitem [{mat()}]{matlab}%
  \BibitemOpen
  \href@noop {} {}\bibinfo {note} {Https://www.mathworks.com/}\BibitemShut
  {NoStop}%
\bibitem [{\citenamefont {Fente}\ \emph {et~al.}(2016)\citenamefont {Fente},
  \citenamefont {Herrera}, \citenamefont {Guillam\'on}, \citenamefont
  {Suderow}, \citenamefont {Ma\~nas Valero}, \citenamefont {Galbiati},
  \citenamefont {Coronado},\ and\ \citenamefont {Kogan}}]{PhysRevB.94.014517}%
  \BibitemOpen
  \bibfield  {author} {\bibinfo {author} {\bibfnamefont {A.}~\bibnamefont
  {Fente}}, \bibinfo {author} {\bibfnamefont {E.}~\bibnamefont {Herrera}},
  \bibinfo {author} {\bibfnamefont {I.}~\bibnamefont {Guillam\'on}}, \bibinfo
  {author} {\bibfnamefont {H.}~\bibnamefont {Suderow}}, \bibinfo {author}
  {\bibfnamefont {S.}~\bibnamefont {Ma\~nas Valero}}, \bibinfo {author}
  {\bibfnamefont {M.}~\bibnamefont {Galbiati}}, \bibinfo {author}
  {\bibfnamefont {E.}~\bibnamefont {Coronado}}, \ and\ \bibinfo {author}
  {\bibfnamefont {V.~G.}\ \bibnamefont {Kogan}},\ }\bibfield  {title} {\emph
  {\bibinfo {title} {Field dependence of the vortex core size probed by
  scanning tunneling microscopy}},\ }\href {\doibase
  10.1103/PhysRevB.94.014517} {\bibfield  {journal} {\bibinfo  {journal} {Phys.
  Rev. B}\ }\textbf {\bibinfo {volume} {94}},\ \bibinfo {pages} {014517}
  (\bibinfo {year} {2016})}\BibitemShut {NoStop}%
\bibitem [{\citenamefont {A.~Fente}()}]{FentePhD}%
  \BibitemOpen
  \bibfield  {author} {\bibinfo {author} {\bibfnamefont {P.}~\bibnamefont
  {A.~Fente}},\ }\href@noop {} {}\bibinfo {note}
  {Http://hdl.handle.net/10486/678930}\BibitemShut {NoStop}%
\end{thebibliography}
%
\end{document}